\documentclass[aps,pra,superscriptaddress,twocolumn,floatfix]{revtex4-2}

\usepackage{comment}
\usepackage{lipsum}
\usepackage[utf8]{inputenc}
\usepackage{amssymb}
\usepackage{amsmath}
\usepackage{amsfonts}
\usepackage{bm}
\usepackage{mathtools}
\usepackage{bbold}
\usepackage{upgreek}
\usepackage{xfrac}
\usepackage{soul}
\usepackage{bm}
\usepackage{comment}
\usepackage{xcolor}
\usepackage{mathtools}
\usepackage{qcircuit}
\usepackage{rotating}
\usepackage{xspace}
\usepackage{placeins}
\usepackage{booktabs}
\usepackage{etoolbox}

\usepackage{natbib}

\usepackage{pgfplots}
\pgfplotsset{width=10cm,compat=1.9}

\usepackage[colorlinks=true,citecolor=blue]{hyperref}
\hypersetup{colorlinks=true,citecolor=blue,linkcolor=blue,urlcolor=blue}
\usepackage{url}

\usepackage{csvsimple}
\usepackage{tabularx}
\usepackage{multirow}
\usepackage{xstring} 
\usepackage{longtable}
\usepackage{braket}

\newcolumntype{C}{>{$}c<{$}}

\AtBeginDocument{%
  \heavyrulewidth=.08em
  \lightrulewidth=.05em
  \cmidrulewidth=.03em
  \belowrulesep=.65ex
  \belowbottomsep=0pt
  \aboverulesep=.4ex
  \abovetopsep=0pt
  \cmidrulesep=\doublerulesep
  \cmidrulekern=.5em
  \defaultaddspace=.5em
}

\usepackage{txfonts}
\DeclareSymbolFont{matha}{OML}{txmi}{m}{it}
\DeclareMathSymbol{\varv}{\mathord}{matha}{118}
\usepackage{bbm}
\usepackage{subfiles} 
\graphicspath{{images/}{../images/}}%

\newcommand{\Gb}{\mathcal{G}b}
\newcommand{\SI}{Supplementary Information\xspace}
\newcommand{\IBMQ}{\emph{IBMQ-27 Ehningen}\xspace}

\DeclareMathOperator{\Prob}{P}
\DeclareMathOperator{\ReLU}{ReLU}

\begin{document}

\title{An Amplitude-Based Implementation of the Unit Step Function on a Quantum Computer}

\author{Jonas Koppe}
\email[Correspondence: Jonas Koppe ]{(jonas.koppe@itwm.fraunhofer.de)}
\affiliation{Department of Financial Mathematics, Fraunhofer ITWM}

\author{Mark-Oliver Wolf}
\affiliation{Department of Financial Mathematics, Fraunhofer ITWM}

\begin{abstract}
    \noindent Modelling non-linear activation functions on quantum computers is vital for quantum neurons employed in fully quantum neural networks, however, remains a challenging task. We introduce an amplitude-based implementation for approximating non-linearity in the form of the unit step function on a quantum computer. Our approach expands upon repeat-until-success protocols, suggesting a modification that requires a single measurement only. We describe two distinct circuit types which receive their input either directly from a classical computer, or as a quantum state when embedded in a more advanced quantum algorithm. All quantum circuits are theoretically evaluated using numerical simulation and executed on Noisy Intermediate-Scale Quantum hardware. We demonstrate that reliable experimental data with high precision can be obtained from our quantum circuits involving up to 8 qubits, and up to 25 CX-gate applications, enabled by state-of-the-art hardware-optimization techniques and measurement error mitigation.
\end{abstract}

\maketitle

\patchcmd{\section}
    {\centering}{\raggedright}{}{}
 \patchcmd{\subsection}
    {\centering}{\raggedright}{}{}

\section*{Introduction}

\noindent Identifying quantum algorithms that allow quantum speedups for machine learning is a research area of rising interest \cite{biamonte_quantum_2017}. The integration of artificial neural networks with quantum computation is typically referred to as quantum neural networks \cite{kak_quantum_1995, ezhov_quantum_2000, schuld_quest_2014, wan_quantum_2017, ciliberto_quantum_2018}. A plethora of different construction strategies for quantum neural networks has been reported over the last decade, a comprehensive review \cite{zhao_review_2021} of which however is far beyond the scope of this article. A rough categorization can be made between hybrid approaches that rely on variational quantum algorithms \cite{mcclean_theory_2016, steinbrecher_quantum_2019,  skolik_layerwise_2021, cerezo_variational_2021, abbas_power_2021}, and fully quantum implementations. For the latter, a key ingredient is the development of a quantum version for the Rosenblatt perceptron, involving the calculation of a tensor product, based on which subsequently a non-linear activation occurs \cite{Rosenblatt_1957_6098}. While tensor calculus has been performed on quantum computers \cite{aaronson_read_2015, Daskin2018ASQ, tacchino_artificial_2019}, the implementation of non-linear activation functions remains a challenging task \cite{cao2017quantum, hu_2018_realQuNeuron, Torrontegui_2019, yan_nonlinear_2020, chakraborty_analytical_2020, allcock_quantum_2020, tacchino_quantum_2020, miller_quantum_2021, maronese_quantum_2022}.

In principle, non-linearities can be modeled using the Unit Step Function (USF), returning zero for negative function inputs, and one otherwise. The USF is also known as the Heaviside (step) function or indicator function \cite{bracewell_heaviside_2000}, originally developed in operational calculus, or the positive part of a function, commonly used in Fourier analysis \cite{savare_positivePart_1996} and finance \cite{korn_2010_monte}. Recently, different approaches to implement non-linearity on a quantum computer have been reported, involving phase encoding and inverse Quantum Fourier Transform \cite{yan_nonlinear_2020}, Taylor expansion \cite{maronese_quantum_2022}, as well as bit-wise comparison \cite{rebentrost_quantum_2018, WoernerEgger2019, vazquez_efficient_2020, Stamatopoulos2020, chakrabarti_threshold_2021}. Moreover, it has been known \cite{cao2017quantum, hu_2018_realQuNeuron, Torrontegui_2019} that specifically the USF can in theory be implemented on quantum computers using repeat-until-success protocols \cite{wiebe_floating_2013, paetznick_2014_RepeatuntilsuccessND, Bocharov_2015_EfficientSO, wiebe_arithmetic_2016, Guerreschi_2019_fixedpointAE}.
However, to the best of our knowledge, a corresponding implementation has not been reported yet.

In this article, we suggest a novel approach that modifies the repeat-until-success gearbox circuit \cite{wiebe_floating_2013, wiebe_arithmetic_2016} to avoid mid-circuit measurements, allowing to encode an arbitrarily close approximation of the USF on the amplitude of qubits.
Specifically, we construct a quantum circuit performing a unitary transformation that, based on a given input, prepares a quantum register to represent the corresponding output of the USF.
It is generally distinguished between passing the input to the circuit either (A) directly as a floating-point value from a classical computer, referred to as classical input, or (B) via another quantum state, referred to as quantum-state input. For (A), we first demonstrate the basic implementation for the USF, and subsequently augment the corresponding circuits for computing other non-linear activation functions, e.g., the Rectified Linear Unit (ReLU) \cite{nair_relu_2010}. For (B), we analyze quantum-state inputs representing a single input angle. Furthermore, we extend our analysis to passing on a quantum state in superposition to the suggested circuit, i.e., for computing the average value of the USF over a set of inputs.

The performance of all quantum circuits shown in this article is not only evaluated by numerical simulation assuming an ideal, noise-free quantum computer, but also tested on the IBM Quantum device in Ehningen, Germany. For implementing quantum circuits on Noisy Intermediate-Scale Quantum (NISQ) hardware \cite{preskill_2018_quantumcomputingin}, we found the number of employed CX gates to be the primary criterion for obtaining reliable results. Therefore, in the following we will restrict the characterization of quantum circuits on NISQ devices to the number of required CX-gate applications.

\section*{Theoretical Considerations}

\begin{figure}
	\centering
	\includegraphics[width=0.9\columnwidth]{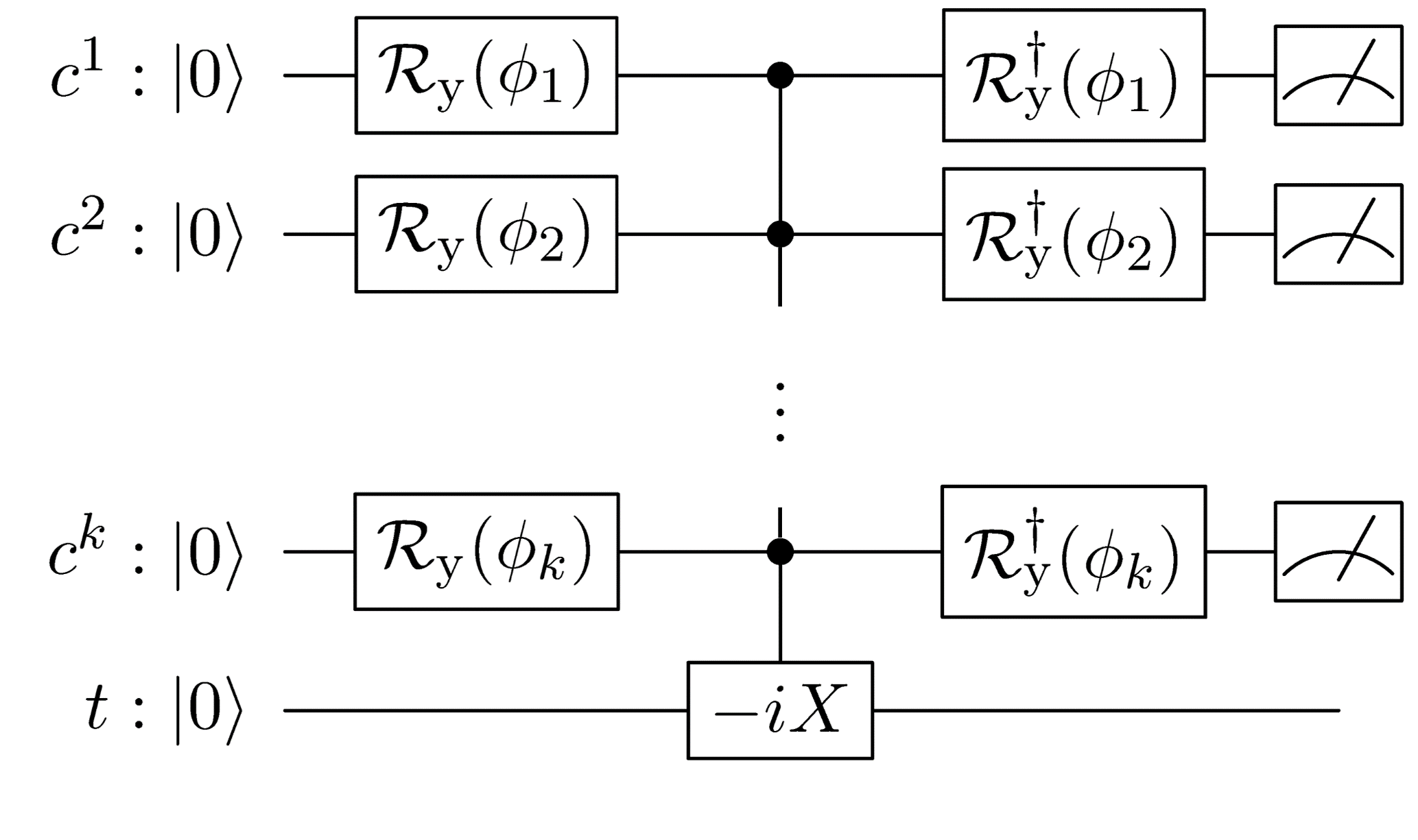}
	\caption{
		A version of the gearbox circuit $C(\pmb{\phi})$ suggested by Wiebe et al \cite{wiebe_floating_2013, wiebe_arithmetic_2016}.
	}
	\label{fig:wiebe_circuit}
\end{figure}

\begin{figure*}[tb]
    \centering
    \includegraphics[width=\textwidth]{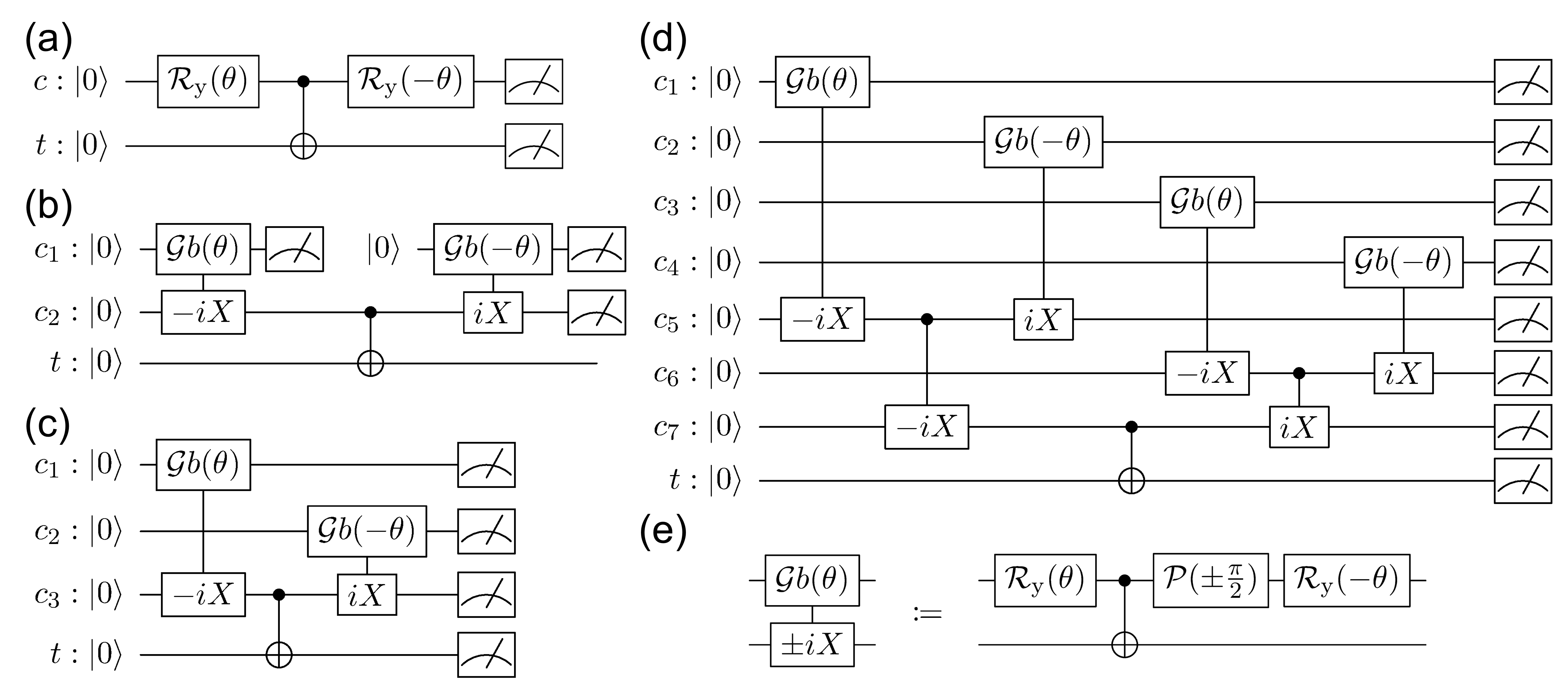}
    \caption{
    Quantum circuits implementing the USF for different levels of approximation $d$.
    (a) Single-step gearbox with $d=1$,
    (b) double-step gearbox with $d=2$ according to repeat-until-success protocols \cite{wiebe_floating_2013},
    (c) double-step gearbox with $d=2$, and
    (d) triple-step gearbox with $d=3$ as suggested in this article.
    (e) Definition of the elementary gearbox-circuit element used in (b), (c), and (d).
    Numbers of the control qubits are given as subscripts to distinguish them from the control qubits shown in Figure~\ref{fig:wiebe_circuit}.
    }
    \label{fig:our_gearbox_circuits}
\end{figure*}

%
%
\noindent In order to synthesize an arbitrarily small rotation, Wiebe et al. proposed the so-called gearbox circuit $C(\pmb{\phi})$ demonstrated in Figure~\ref{fig:wiebe_circuit} \cite{wiebe_floating_2013, wiebe_arithmetic_2016}. The name stems from the fact that the $k$ input angles $\pmb{\phi} = \phi_1,\dots,\phi_k$ applied to the $k$ control qubits $c^1,\dots,c^k$ are used to produce a rotation involving a much finer rotation angle $\theta$ applied to the target qubit, where $\sin^2(\theta) = \sin^2(\phi_1) \cdots \sin^2(\phi_k)$. The entire transformation can formally be written as
\begin{equation}\label{eq:basic_gearbox}
\begin{split}
    C(\pmb{\phi}) \ket{0^{\otimes k}}_c \ket{0}_t
    =
    \; &\rho (\theta) \ket{0^{\otimes k}}_c \; e^{ -i \arctan \left[  \tan^2 \theta \right] X } \ket{0}_t\\
    &- \sqrt{ 1 - \rho^2(\theta) } \ket{\emptyset^{\otimes k}}_c \; e^{ -i \frac{\pi}{4} X } \ket{0}_t,
\end{split}
\end{equation}
where we have used the indices $c$ and $t$ to indicate the control (qubit) register and target qubit, respectively. $\ket{\emptyset^{\otimes k}}_c$ is a compact notation representing the sum over all possible states of the control register except for the all-zero state $\ket{0^{\otimes k}}_c$. Notably, a rotation involving $\theta$, i.e., the transformation
$ \ket{0}_t \mapsto e^{ -i \arctan \left[ \tan^2 \theta \right] X } \ket{0}_t $,
is successfully applied to the target qubit only if the control register is in $\ket{0^{\otimes k}}_c$. The corresponding success probability is given by
\begin{equation}\label{eq:success_prob}
	\rho^2(\theta) \coloneqq \sin^{4}(\theta) + \cos^4(\theta).
\end{equation}
Otherwise $C(\pmb{\phi})$ transforms the target qubit according to $\ket{0}_t \mapsto e^{ -i \frac{\pi}{4} X } \ket{0}_t$. Without exactly determining the state of the target qubit after circuit execution, measuring the control register consequently allows to ascertain which transformation has been performed on the target qubit. Repeating the circuit $C(\pmb{\phi})$ until the state $\ket{0^{\otimes k}}_c$ is measured in turn ensures that the desired transformation
$ \ket{0}_t \mapsto e^{ -i \arctan \left[  \tan^2 \theta \right] X } \ket{0}_t $
has occurred.
This is known as the repeat-until-success principle \cite{wiebe_arithmetic_2016, cao2017quantum, paetznick_2014_RepeatuntilsuccessND, Bocharov_2015_EfficientSO, Guerreschi_2019_fixedpointAE}. Moreover, Wiebe et al. have theoretically shown that a nested version of the gearbox circuit can be built up recursively, i.e., the transformation on the target qubit from the gearbox circuit of Figure~\ref{fig:wiebe_circuit} serves itself as the input of an outer gearbox circuit as shown in Figure~\ref{fig:our_gearbox_circuits}b.
The transformation performed on the outermost target qubit of a $d$-times nested gearbox circuit given all control qubits are found in $\ket{0}_c$ is then given as
$ \ket{0}_t \mapsto e^{ -i \arctan \left[ \tan^{2^d} \theta \right] X } \ket{0}_t $ \cite{wiebe_floating_2013, wiebe_arithmetic_2016}. It is emphasized that generally the success probability $\rho^2(\theta)$ depends on $d$, as detailed in the \SI (Section~SI~1). However, in the following only the case for $d=1$ as given in Eq.~\eqref{eq:success_prob} will be relevant.
\begin{figure}
	\centering
	\includegraphics[width=0.9\columnwidth]{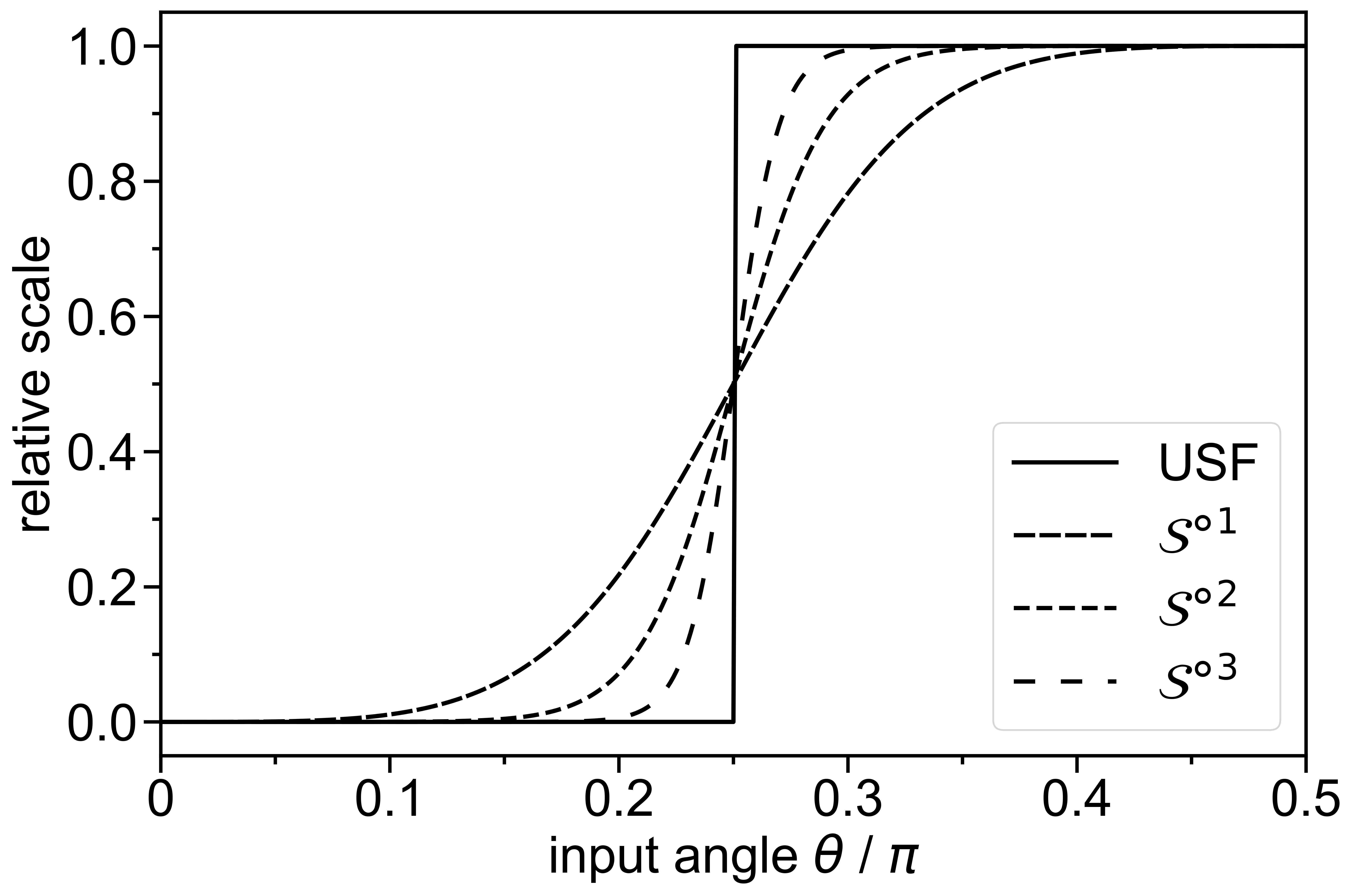}
	\caption{
		USF and its approximations $\mathcal{S}^{\circ d} (\theta)$ (see Eq.~\eqref{eq:our_approx}) for different levels $d = 1,2$, and $3$ on the input space used in this article.
	}
	\label{fig:sod}
\end{figure}

\section*{Results}

For implementing an approximation of the USF we use the square-wave property of the gearbox \cite{wiebe_arithmetic_2016}. For our purpose, the most basic gearbox version is sufficient, involving only a single control qubit, and likewise a single input angle $\phi_1=\theta$, as shown in Figure~\ref{fig:our_gearbox_circuits}a. In the following, this is referred to as a single-step gearbox $\Gb(\theta)$ (also cf.\ Figure~\ref{fig:our_gearbox_circuits}e). The transformation from Eq.~\eqref{eq:basic_gearbox} can then be expressed as
\begin{equation}\label{eq:our_single_gb}
\begin{split}
    &\Gb (\theta) \ket{0}_c \ket{0}_t \\
    &=
    \rho (\theta) \ket{0}_c \;
    \Big(
    \cos \left[ \arctan [ \tan^2 \theta ] \right] \ket{0}_t \\
    &\hspace{1.6cm} +
    \sin \left[ \arctan [ \tan^2 \theta ] \right] \ket{1}_t
    \Big) \\
    &\hspace{.3cm} - \sin (\theta) \cos (\theta) \ket{1}_c \; \big( \ket{0}_t + \ket{1}_t \big).
\end{split}
\end{equation}
Assuming the control qubit to be in the zero state $\ket{0}_c$, the probability of finding the target qubit in $\ket{1}_t$ is given by $\mathcal{S}^{\circ 1} (\theta) \coloneqq \sin^2 \left[ \arctan [ \tan^2 \theta ] \right]$; the corresponding trajectory is demonstrated in Figure~\ref{fig:sod}.
Clearly $\mathcal{S}^{\circ 1}(\theta)$ is reminiscent of a sigmoid function \cite{han_1995_sigmoid} for $\theta \in [0,\pi/2]$, and thus an appropriate approximation for the USF. Extending this observation to a $d$-nested gearbox circuit, in the following referred to as a $d$-step gearbox, the probability of finding the target qubit in $\ket{1}_t\,$, provided all control qubits have prior been found in $\ket{0}_c\,$, is accordingly given by
\begin{equation}\label{eq:our_approx}
	\mathcal{S}^{\circ d} (\theta) \coloneqq \sin^2 \left[ \arctan [ \tan^{2^d} \theta ] \right]. 
\end{equation}%

\noindent Trajectories for $d=2$ and $3$ are likewise shown in Figure~\ref{fig:sod}. Even though it is theoretically known that in such a way an approximation of the USF is encoded in the amplitude of the target qubit \cite{wiebe_arithmetic_2016, cao2017quantum}, to the best of our knowledge, an exact protocol for exploiting this fact has not been reported yet. Over the course of this article, we will introduce two alternatives for retrieving $\mathcal{S}^{\circ d}(\theta)$ from the quantum states produced by the $d$-step gearbox circuits. The application of both is demonstrated for the two scenarios described above, i.e., (A) a classical input, where the input for the gearbox circuit is a single angle $\theta \in [0, \pi/2]$ directly passed on by a classical computer, based on which the target qubit is ideally transformed according to
\begin{equation}\label{eq:quantum_usf}
   \ket{0}_t \longmapsto
    \begin{cases}
        \ket{0}_t \text{ for } \theta \leq \frac{\pi}{4}, \\
        \ket{1}_t \text{ for } \theta > \frac{\pi}{4},
    \end{cases}
\end{equation}
\noindent as demonstrated in Figure~\ref{fig:sod} (USF). For the second scenario (B), the task is extended to passing a quantum-state input to the gearbox circuit, where each eigenstate in the measurement basis represents a different input angle $\theta_j \in [0,\pi/2]$ used for the transformation in Eq.~\eqref{eq:quantum_usf}.

\subsection*{(A) Classical Input}

\noindent \textbf{Basic USF Version.} Consider the single-step gearbox circuit from Figure~\ref{fig:our_gearbox_circuits}a, where we wish the measurement to reflect $\mathcal{S}^{\circ 1}(\theta)$. Evaluation of Eq.~\eqref{eq:our_single_gb} however shows that measuring the probability for the state $\ket{1}_t \ket{0}_c$ yields $\rho^2(\theta) \mathcal{S}^{\circ 1}(\theta)$. It is indeed possible to remove the distortion $\rho^2(\theta)$ by combining the probabilities measured for both states involving $\ket{0}_c\;$: the probability of measuring the target qubit in state $\ket{1}_t$ under the condition that the control qubit is found in state $\ket{0}_c$ is given by
\begin{widetext}
\begin{equation}\label{eq:cond_prob_measurement}
\begin{split}
    \Omega ( 1 \mid 0 )
    &\coloneqq
    \Prob \left( \ket{1}_t \mid \ket{0}_c \right)
    =
    \frac{
    \left| \left( \bra{1}_t \bra{0}_c \right) \ket{\tau_{\Gb}} \right|^2
    }{
    \left| \left( \bra{0}_t \bra{0}_c \right) \ket{\tau_{\Gb}} \right|^2
    +
    \left| \left( \bra{1}_t \bra{0}_c \right) \ket{\tau_{\Gb}} \right|^2
    } \\
    &=
    \frac{
    \rho^2(\theta) \sin^2 \left[ \arctan [ \tan^2 \theta ] \right]
    }{
    \rho^2(\theta) \cos^2 \left[ \arctan [ \tan^2 \theta ] \right]
    +
    \rho^2(\theta) \sin^2 \left[ \arctan [ \tan^2 \theta ] \right]
    }
    =
    \sin^2 \left[ \arctan [ \tan^2 \theta ] \right]
    = \mathcal{S}^{\circ 1}(\theta),
\end{split}
\end{equation}
\end{widetext}

\noindent where $\ket{\tau_{\Gb}}$ represents the state of both gearbox qubits before initiating the measurement. Unfortunately, this evaluation cannot simply be extended to $d$-step gearbox circuits and thus, to higher levels of approximation $d$ for the USF, as they involve repeat-until-success protocols that rely on mid-circuit measurements \cite{paetznick_2014_RepeatuntilsuccessND, Bocharov_2015_EfficientSO, cao2017quantum,Guerreschi_2019_fixedpointAE, wiebe_floating_2013, wiebe_arithmetic_2016} (cf.\ Figure~\ref{fig:our_gearbox_circuits}b).
\begin{figure*}
    \centering
    \includegraphics[width=\textwidth]{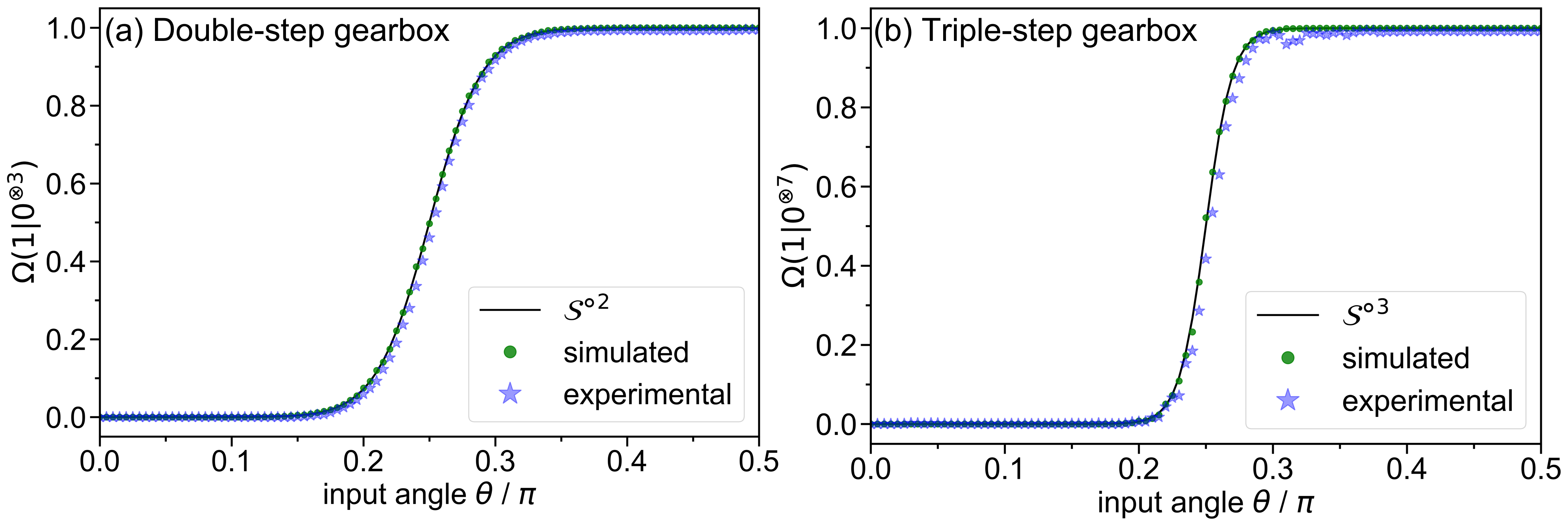}
    \caption{
    Results for the implementation of the USF at approximation levels $d = 2$ and $3$ in (a) and (b), respectively. Analytical curves are based on Eq.~\eqref{eq:our_approx}. Simulated data (green circles) was obtained by executing the quantum circuits shown in Figure~\ref{fig:our_gearbox_circuits}(c) and (d). For the experimental data (blue stars), the respective circuits have been optimized and executed on \IBMQ. Optimum quantum circuits are demonstrated in Figure~S4. For the simulated and experimental circuit runs, the input space $\theta \in [0,\pi/2]$ has been covered by $101$ increments $\theta_j = j \pi/200, j=0,\dots,100$, where for each data point $10^5$ circuit executions were conducted.
    }
    \label{fig:double_and_triple_gearbox}
\end{figure*}
\noindent We propose an alternative implementation by allocating each gearbox element within a nested circuit to a new set of qubits, as demonstrated in the circuit modification in Figure~\ref{fig:our_gearbox_circuits}c for the double-step gearbox. In Figure~\ref{fig:our_gearbox_circuits}d, the corresponding extension of the triple-step gearbox is shown. This implementation principally requires $2^d$ measurable qubits, and the application of $2^d-1$ CX gates. The corresponding generalization of Eq.~\eqref{eq:cond_prob_measurement} is given by
\begin{equation}\label{eq:d-step_cond_prob_measurement}
    \Omega ( 1 \mid 0^{ \otimes 2^d-1 } )
    =
    \sin^2 \left[ \arctan [ \tan^{2^d} \theta ] \right]
    =
    \mathcal{S}^{\circ d}(\theta).
\end{equation}
For a more thorough mathematical treatment, the reader is referred to the \SI (Section~SI~1). To demonstrate the validity of this approach, we performed simulations of the quantum circuits shown in Figure~\ref{fig:our_gearbox_circuits}c and d, implementing approximation level $d=2$ (double-step gearbox) and $d=3$ (triple-step gearbox) for the unit step function, respectively. 
An ideal, noise-free quantum computer with all-to-all connectivity was assumed, requiring $4$ qubits and $3$ CX gates for $d=2$, and $8$ qubits and $7$ CX gates for $d=3$.
The input space $\theta \in [0,\pi/2]$ was covered by 101 increments, where each data point represent an estimate for $\Omega ( 1 \mid 0^{ \otimes 3 } )$ and $\Omega ( 1 \mid 0^{ \otimes 7 } )$, obtained from $10^5$ circuit executions.
The results (green circles) are compared to $\mathcal{S}^{\circ 2}(\theta)$ and $\mathcal{S}^{\circ 3}(\theta)$ in Figure~\ref{fig:double_and_triple_gearbox}. Clearly, the numerical simulations are in excellent agreement with their respective analytical form, confirming Eq.~\eqref{eq:d-step_cond_prob_measurement}. In order to assess the applicability of the implementation on NISQ devices, both quantum circuits were optimized for the 27-qubit IBM Quantum system in Ehningen, Germany (\IBMQ). The hardware-optimization procedure is detailed in the \hyperref[sec:methods]{Methods}~section. It should be emphasized here that due to the restricted connectivity based on the heavy-hexagon qubit architecture, state-swapping was required for the triple-step gearbox circuit, resulting in an overhead of $4$ CX gates, and thus, in $11$ CX-gate applications in total. The results are likewise shown in Figure~\ref{fig:double_and_triple_gearbox} (blue stars). Again, $10^5$ circuit executions were performed for each input angle $\theta$. In general, the data reflects the simulated behavior with high precision. Yet deviations do occur for input angles at about the step $\theta \approx \pi/4$, where the experimentally obtained values systematically remain below the numerical prediction. Even though this effect becomes more pronounced for the triple-step gearbox, it does not qualitatively affect the approximation of the USF.
%
\begin{figure*}
    \centering
    \includegraphics[width=\textwidth]{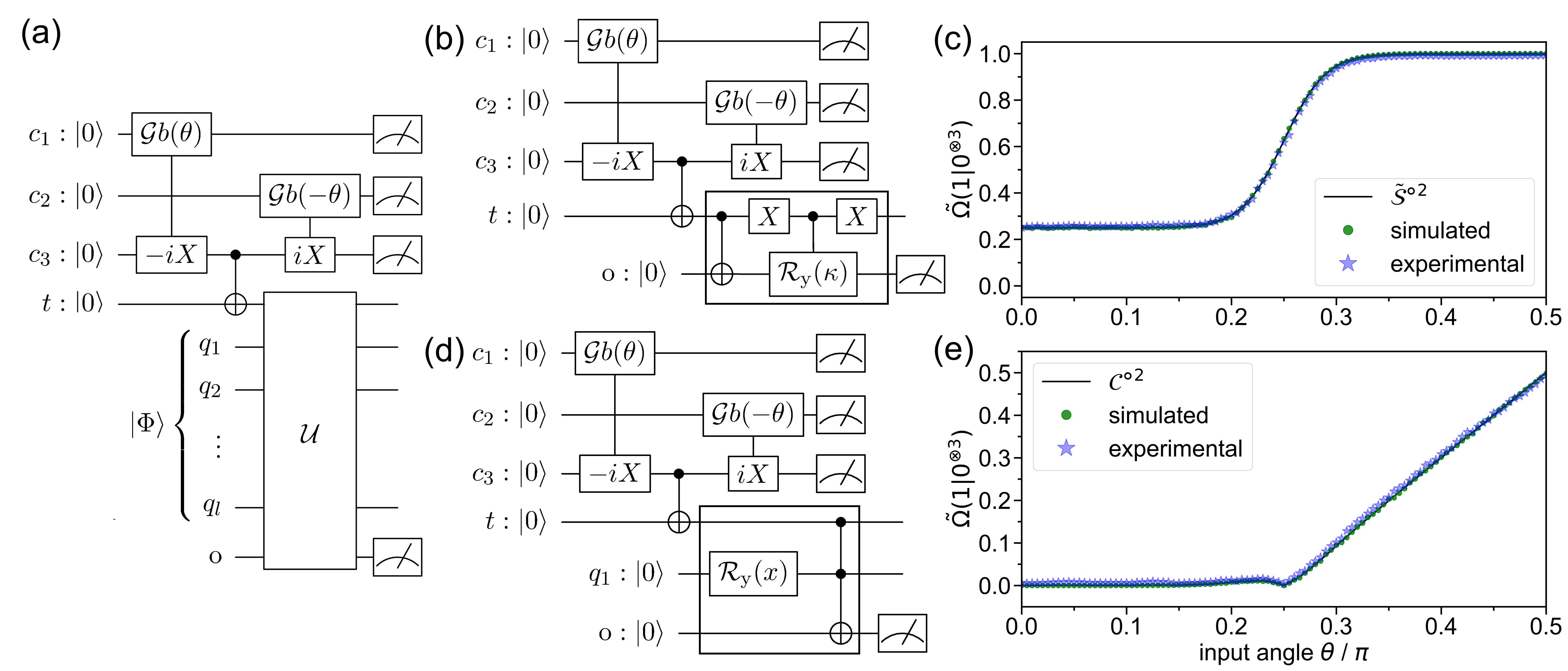}
    \caption{
    (a) Schematic representation for combining an arbitrary unitary transformation $\mathcal{U}$ with the double-step gearbox from Figure~\ref{fig:our_gearbox_circuits}c. (b)/(d) Quantum circuits approximating the transformations given in Eq.~\eqref{eq:primitive_example_U} and Eq.~\eqref{eq:quantum_relu_mapping}, respectively. The black boxes indicate the additional unitary transformations $\mathcal{U}$ as generally introduced in (a). (c)/(e) Analytical, simulated, and experimental data for the quantum circuits from (b) and (d), respectively. All conditions are identical to those reported for Figure~\ref{fig:double_and_triple_gearbox}. Optimum quantum circuits on \IBMQ are shown in Figures~S2 and S3.
    }
    \label{fig:experimental_relu}
\end{figure*}
\newline
\newpage
\noindent \textbf{Variations.} We emphasize that the evaluation strategy described above for retrieving $\mathcal{S}^{\circ d}(\theta)$ can be incorporated into more advanced quantum circuits. For instance, consider a unitary $\mathcal{U}$ requiring as an input the (approximated) USF $\mathcal{S}^{\circ d}(\theta)$, which performs a transformation on an additional input state $\ket{\Phi}$, and maps the result to an output qubit $o$. A schematic representation of this type of circuit involving the double-step gearbox is suggested in Figure~\ref{fig:experimental_relu}a. It is important to highlight that only the state of the target qubit $t$ must be passed to $\mathcal{U}$. Eventually, data evaluation is done analogously to Eq.~\eqref{eq:d-step_cond_prob_measurement}, however replacing the target qubit $t$ with the output qubit $o$, i.e.,
\begin{equation}\label{eq:relu_circuit_measurement}
    \tilde{\Omega} \left( 1 \;\middle|\; 0^{ \otimes 2^d-1 } \right)
    \coloneqq
    \Prob \left( \ket{1}_o \;\middle|\; \ket{0^{ \otimes 2^d-1 }}_c \right).
\end{equation}
As a primitive example for $\mathcal{U}$ requiring no additional input $\ket{\Phi}$, assume we wish to modify the transformation given in Eq.~\eqref{eq:quantum_usf} and instead encode
\begin{equation}\label{eq:primitive_example_U}
    \ket{0}_o \longmapsto
    \begin{cases}
        \cos \kappa \ket{0}_o
        +
        \sin \kappa \ket{1}_o
        &\text{ for } \theta \leq \frac{\pi}{4}, \\
        \ket{1}_o &\text{ for } \theta > \frac{\pi}{4},
    \end{cases}
\end{equation}
on the output qubit $o$. This effectively allows shifting the first plateau from $0$ to $\sin^2 \kappa$. In Figure~\ref{fig:experimental_relu}b we demonstrate a quantum circuit approximating this transformation using a double-step gearbox, such that we expect to observe the function
$ \tilde{\mathcal{S}}^{\circ 2}(\theta) \coloneqq \sin^2(\theta) \kappa \left( 1 - \mathcal{S}^{\circ 2}(\theta)\right) + \mathcal{S}^{\circ 2}(\theta)$. Another important transformation that can be generated from the USF is the so-called $\ReLU$ activation function \cite{nair_relu_2010}, commonly defined as $\ReLU(x) \coloneqq \max(0,x)$. Here, the output of the USF is multiplied with the identity $f(x) = x$. Note that the unit step must necessarily coincide with the root of $f(x)$. For the implementation on a quantum computer this requires synchronization of the gearbox input $\theta$ and the function input $x$.
A variety of strategies could be used for this purpose.
For simplicity, here we rigorously encode $f(x)$ on a single input qubit $q_1$ using an $\mathcal{R}_{\mathrm{y}}(x)$ gate with
\begin{equation}
    x = x(\theta)
    = \arcsin \left( \sqrt{ \left| \frac{2\theta}{\pi} - \frac{1}{2} \right| } \right),
\end{equation}
where we make use of the fact that for all $x \leq \pi/4$ the corresponding gearbox output $\mathcal{S}^{\circ d} (\theta)$ evaluates to zero. We thereby guarantee that $f(x\,(\theta))\in[0, 1/2]$ and $f( x\,(\pi/4) ) = 0$. The entire $\ReLU$ transformation, formally written as
\begin{equation}\label{eq:quantum_relu_mapping}
    \ket{0}_o \longmapsto
    \begin{cases}
        \ket{0}_o &\text{ for } \theta \leq \frac{\pi}{4}, \\
        \sqrt{f(x\,(\theta))} \ket{1}_o &\text{ for } \theta > \frac{\pi}{4},
    \end{cases}
\end{equation}
is approximated by the quantum circuit demonstrated in Figure~\ref{fig:experimental_relu}d, employing a double-step gearbox. The expected output is then given by $\mathcal{C}^{\circ 2}(\theta) \coloneqq \mathcal{S}^{\circ 2}(\theta) \cdot f(x\,(\theta))$. Using identical conditions as described in Figure~\ref{fig:double_and_triple_gearbox}, we tested the performance of the quantum circuit from Figure~\ref{fig:experimental_relu}b for $\kappa = \arcsin \sqrt{1/4}$, i.e., expecting a first plateau at $1/4$, and the quantum circuit from Figure~\ref{fig:experimental_relu}d on an ideal, noise-free quantum computer as well as on \IBMQ. The corresponding analytical, simulated and experimental data is demonstrated in Figure~\ref{fig:experimental_relu}c and e, respectively. As observed before, the simulated but also experimental data is in excellent agreement with the analytically expected behavior.

\subsection*{(B) Quantum-State Input}

\begin{figure}
    \centering
    \includegraphics[width=1.\columnwidth]{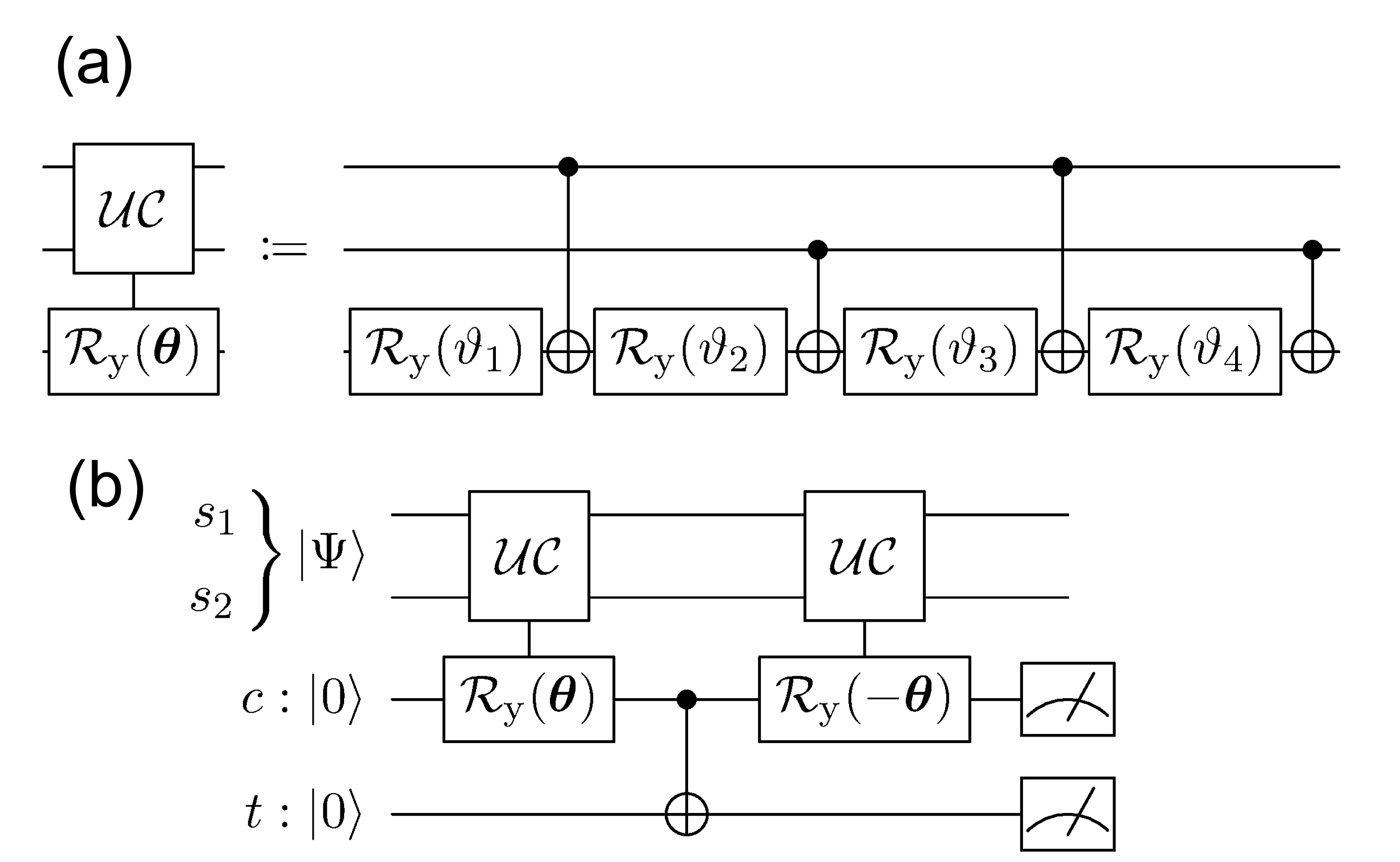}
    \caption{
    (a) Uniformly-controlled rotations involving two state qubits. For eventually implementing the correct rotation according to $\theta_j/\pi \in \{ 0.15,0.2,0.4,0.45 \}$, angles must be converted as described in \cite{mottonen_2004_generalMQGates}. The corresponding angles here are given by $\vartheta_j/\pi \in \{ 0.3,-0.025,0,-0.125 \}$. (b) Uniformly-controlled rotations as shown in (a) embedded in the single-step gearbox. The state register is represented by $\ket{\Psi} = \sum_j \alpha_j \ket{\psi_j}$.
    }
    \label{fig:uniformly-controlled_implementation}
\end{figure}

\noindent The results discussed in the previous section generally demonstrate the ability to approximate the USF on a quantum computer. However, directly passing an input angle $\theta$ to the gearbox circuit requires communication with a classical computer. For many future applications we rather assume gearbox circuits to receive a quantum state $\ket{\Psi}$ as an input, with each eigenstate of the measurement basis $\ket{\psi_j}$ representing a different input angle $\theta_j$. In this section, we therefore demonstrate results for a single-step gearbox capable of considering four arbitrarily chosen input angles $\pmb{\theta}/\pi \coloneqq \left(0.15, 0.2, 0.4, 0.45\right)^T$, where each entry is denoted as $\theta_j/\pi, j=0,\dots,3$. These angles are represented by a two-qubit state register $s$, i.e., by the amplitudes of $\ket{\psi_j} \in \left\{ \ket{00}, \ket{01}, \ket{10}, \ket{11} \right\}$, respectively. An efficient approach for passing input angles to the gearbox circuit based on the state of $s$ are uniformly-controlled rotations \cite{mottonen_2004_generalMQGates, shende_2006_synthQLogicCircs}. This generally requires $2^p$ CX-gate applications, where $p$ indicates the number of qubits in the state register. The corresponding circuit element for a two-qubit state register is shown in Figure~\ref{fig:uniformly-controlled_implementation}a. Note that the angles $\vartheta_j$ involved in the depicted sequence of rotations can be obtained from the $\theta_j$, the exact conversion is detailed in \cite{mottonen_2004_generalMQGates}. The incorporation of uniformly-controlled rotations into the single-step gearbox, as demonstrated in Figure~\ref{fig:uniformly-controlled_implementation}b, generally requires $9$ CX gates.
\newline

\begin{figure}
    \centering
    \includegraphics[width=1.\columnwidth]{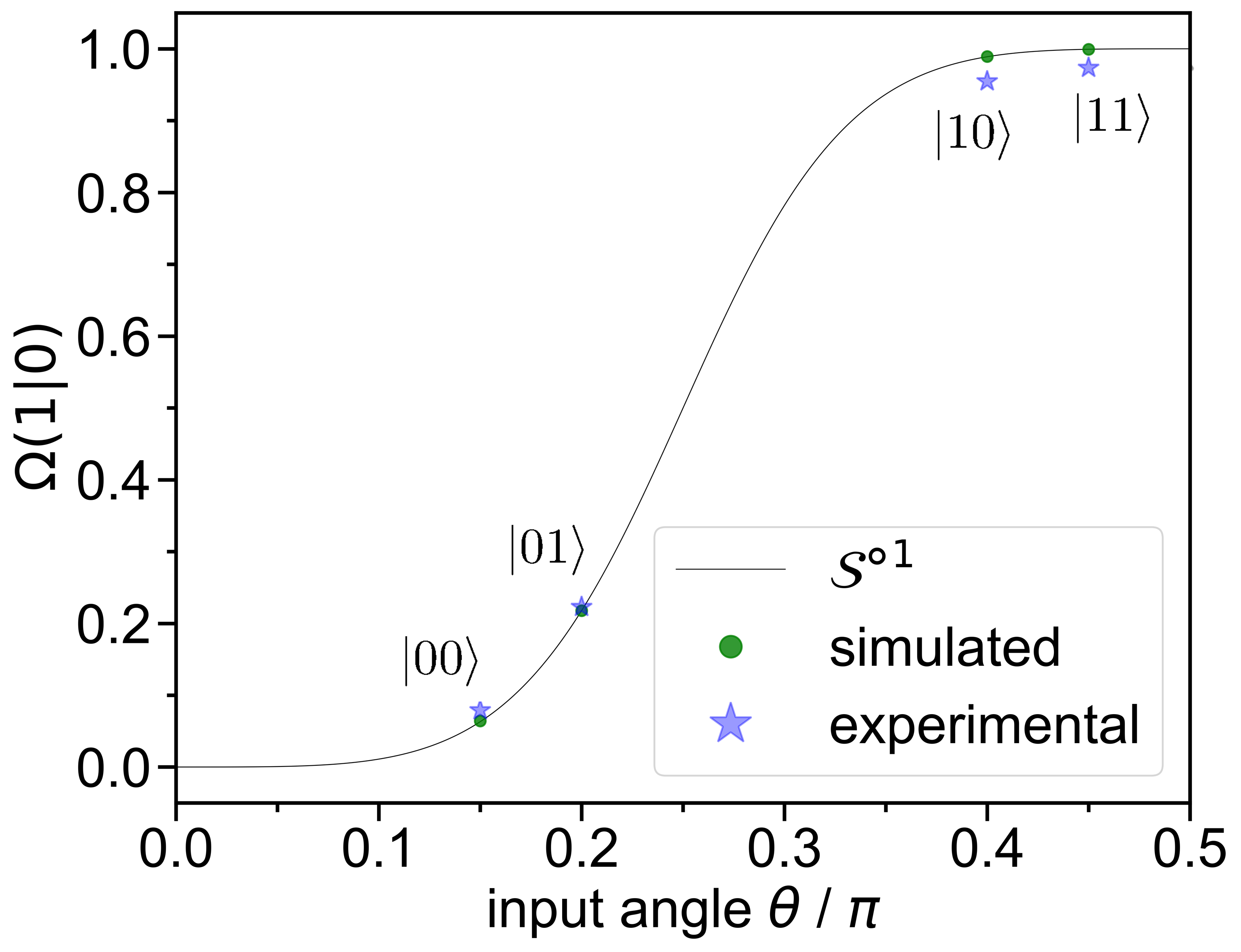}
    \caption{
    Results for the quantum-state input passed to the single-step gearbox as shown in Figure~\ref{fig:uniformly-controlled_implementation}b. The state register is represented by $\ket{\Psi} \in \left\{ \ket{00}, \ket{01}, \ket{10}, \ket{11} \right\}$ as indicated, corresponding to the input angles $\theta_j/\pi \in \{0.15, 0.2, 0.4, 0.45\}$. The curve for $\mathcal{S}^{\circ 1}(\theta)$ is based on Eq.~\eqref{eq:our_approx}.
    Simulated data, illustrated as green circles, was obtained by executing the quantum circuits shown in Figure~\ref{fig:uniformly-controlled_implementation}b $10^5$ times assuming an ideal, noise-free quantum computer with all-to-all connectivity.
    For the experimental data, the circuit has been optimized for \IBMQ. $100$ experimental hardware runs were performed, each likewise comprising $10^5$ circuit executions. The blue stars represent the respective averages. The optimum quantum circuit for $\ket{00}$ on \IBMQ is shown in Figure~S5a.
    }
    \label{fig:state-dependent_implementation}
\end{figure}

\begin{figure}
    \centering
    \includegraphics[width=0.9\columnwidth]{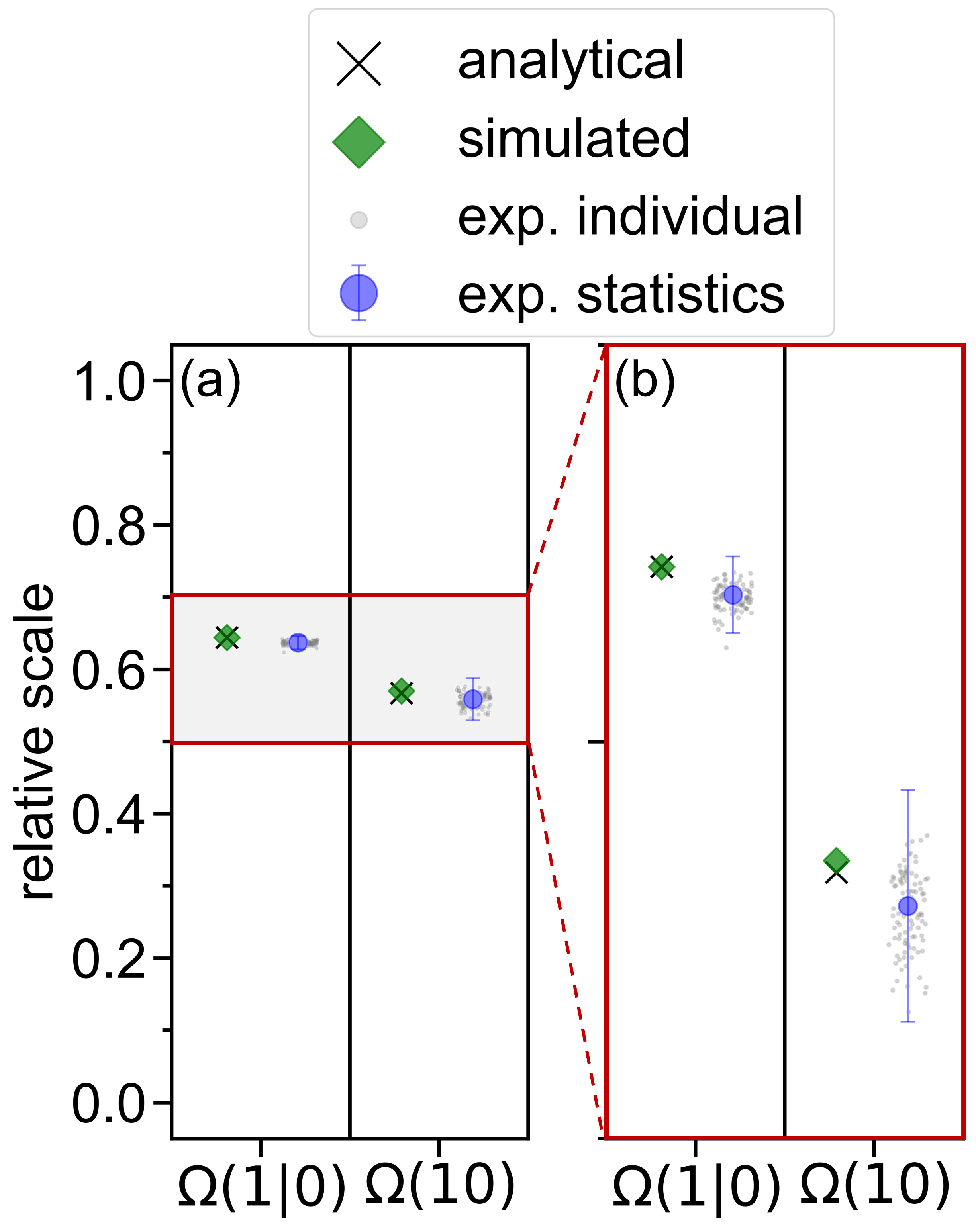}
    \caption{
    (a) Results for the average output of the single-step gearbox from Figure~\ref{fig:uniformly-controlled_implementation}b. (b) Inset indicated with the red box in (a). Analytical values are given by $\Omega(1 \mid 0)\approx0.644$ according to Eq.~\eqref{eq:multi-state_cond_prob}, and $\overline{\mathcal{S}^{\circ 1}}(\pmb{\theta}) = 0.567$ for $\Omega(10)$. Simulated and experimental data for $\Omega(1 \mid 0)$ are based on the circuit from Figure~\ref{fig:uniformly-controlled_implementation}b with the state register represented by $\ket{\Psi} = \ket{+}\ket{+}$. Simulated and experimental data for $\Omega(10)$ stems from executing the subcircuits $\Delta^0, \Delta^2, \Delta^4$ and $\Delta^6$, and post-processing the results as detailed in the main text. For all (sub)circuits, $100$ hardware runs (gray circles) with $10^5$ executions each were conducted. The respective averages and $3\sigma$ levels are indicated by the blue circles and error bars. The optimum quantum circuits on \IBMQ for $\Omega(1 \mid 0)$ and $\Omega(10)$ can be found in Figure~S5. 
    }
    \label{fig:multi-state_implementation_comparison}
\end{figure}

\noindent \textbf{Single Angle.} At first, we consider a situation where the state-qubit register is in one of the eigenstates of the measurement basis, i.e, $\ket{\Psi} = \ket{\psi_j}$, such that only one of the four angles $\theta_j$ is passed to the gearbox circuit. In fact, this is not fundamentally different from the case of a classical input discussed in the previous section; data evaluation can be performed according to Eq.~\eqref{eq:cond_prob_measurement} to obtain $\mathcal{S}^{\circ 1}(\theta_j)$. In Figure~\ref{fig:state-dependent_implementation}, $\mathcal{S}^{\circ 1}(\theta)$ is again shown for the entire input space $\theta\in[0,\pi/2]$.
The results for executing the quantum circuit from Figure~\ref{fig:uniformly-controlled_implementation}b $10^5$ times for each $\ket{\Psi} = \ket{\psi_j}$ assuming an ideal, noise-free quantum computer are represented by the green circles. Additionally, the respective states of the state register are indicated. Clearly, analytical and simulated data is in excellent agreement.
For assessing its performance on a NISQ device, the quantum circuit from Figure~\ref{fig:uniformly-controlled_implementation}b was again optimized for \IBMQ (resulting in 9 CX-gate applications with no overhead), and executed $10^5$ times for each input state. Each of these runs was then repeated $100$ times, from which the resulting average values for $\Omega(1\mid0)$ are indicated by the blue stars in Figure~\ref{fig:state-dependent_implementation}.
%
%
Notably, the experimental data does not achieve the level of precision we expect from the results demonstrated in Figure~\ref{fig:double_and_triple_gearbox}: while for $\theta_j < \pi/4$ estimators for $\Omega(1 \mid 0)$ are found slightly above the expected values, for $\theta_j > \pi/4$ results remain below the analytical and numerical prediction. We assign this to the diminished precision of $\theta_j$ passed to the gearbox. Here $\theta_j$ is generated by the uniformly-controlled rotations, involving $4$ CX-gates and several single-qubit-gate applications, and is thus significantly more prone to error when compared to the classical input $\theta$ from the previous section. Since the individual experimental data ($100$ hardware runs not shown here for clarity) have sufficiently small confidence intervals ($3 \sigma$ levels lie within the size of the symbol), this error seems to be rather systematic than random. Nevertheless, the overall behavior according to $\mathcal{S}^{\circ 1}(\theta)$ is generally well reflected by the experimental results.
\newline\newline
\noindent \textbf{Multiple Angles.} 
A more complex situation occurs when the state register $s$ is no longer in one of the four eigenstates $\ket{\psi_j}$. Here we are particularly concerned with the state qubits in an equal superposition of all four eigenstates, $\ket{\Psi} = \ket{+} \ket{+} = 1/2\sum_{j=0}^{3}\ket{\psi_j}$, i.e., the average output of the gearbox. Unfortunately, the success probability $\rho^2(\theta_j)$ is state-dependent, such that $\overline{\mathcal{S}^{\circ 1}}(\pmb{\theta}) \coloneqq 1/4 \sum_{j=0}^{3} \mathcal{S}^{\circ 1}(\theta_j)$ cannot simply be retrieved by using the evaluation strategy suggested in Eq.~\eqref{eq:cond_prob_measurement}.  Instead, this yields

\begin{equation}\label{eq:multi-state_cond_prob}
    \Omega( 1 \mid 0 )
    =
    \frac{
    \sum_{j=0}^{3} \rho^2(\theta_j) \sin^2 \left[ \arctan [ \tan^2 \theta_j ] \right]
    }{
    \sum_{j=0}^{3} \rho^2(\theta_j)
    },
\end{equation}
where $j$ is the integer representation of the two-bit string in the computational basis. For the chosen input angles $\pmb{\theta}$, Eq.~\eqref{eq:multi-state_cond_prob} evaluates to $\Omega(1 \mid 0) \approx 0.644$, while we wish to find $\overline{\mathcal{S}^{\circ 1}}(\pmb{\theta}) \approx 0.567$. To confirm this bias, we have repeated the procedure described for Figure~\ref{fig:state-dependent_implementation} with the state register given by $\ket{\Psi} = \ket{+}\ket{+}$.
Analytical, simulated, and experimental data is demonstrated in the left panel of Figure~\ref{fig:multi-state_implementation_comparison}a. Here, we additionally showed the individual $100$ runs as small gray circles. Blue circles and error bars indicate the average and $3 \sigma$ levels, respectively.
The inset indicated by the red box is shown in Figure~\ref{fig:multi-state_implementation_comparison}b.
For clarity we shifted analytical and simulated data against the experimental data.
Generally, these results are in agreement with Eq.~\eqref{eq:multi-state_cond_prob}. The experimental estimator for $\Omega(1 \mid 0)$ remains slightly below the predicted value, which however is in accordance with the observations from Figure~\ref{fig:state-dependent_implementation}.

It should be reiterated here, that following Eq.~\eqref{eq:cond_prob_measurement}, for the single-step gearbox the probability of measuring $\ket{0}_c \ket{1}_t$ with a state register in equal superposition is given by
\begin{equation}\label{eq:multi-state_cond_prob_measurement}
    \left| \left( \bra{1}_t \bra{0}_c \right) \ket{\tau_{\Gb}} \right|^2
    =
    \frac{1}{4} \sum_{j=0}^{3} \rho^2(\theta_j) \sin^2\left[ \arctan[ \tan^2 \theta_j ] \right],
\end{equation}
with $\ket{\tau_{\Gb}}$ representing the state for the two involved gearbox qubits. Aiming to obtain $\overline{\mathcal{S}^{\circ 1}}(\pmb{\theta})$ from this result, a transformation must be found performing a state-wise amplitude encoding for $\rho^{-1}(\theta_j)$, which is then appended to the gearbox circuit. Then, a measurement yields
\begin{equation}\label{eq:fixed_multi-state_measurement}
    \Omega(10)
    \coloneqq
    \frac{1}{4} \sum_{j=0}^3 D(\theta_j) \, \rho^2(\theta_j) \sin^2\left[ \arctan[ \tan^2 \theta_j ] \right]
    = \overline{\mathcal{S}^{\circ 1}}(\pmb{\theta}),
\end{equation}

\noindent where $D(\theta_j) \coloneqq \rho^{-2}(\theta)$. Since $\rho^{-1}(\theta_j)$ is a periodic function (cf. Eq.~\eqref{eq:success_prob}), a corresponding transformation may be approximated using its Fourier series, a strategy which has been suggested for similar purposes before \cite{li_universal_2021}. However, due to the structure of the problem, here we chose a different approach than previously reported. Each term from the series expansion is encoded on a different qubit. Eventually, the full transformation can be constructed by adding up all of these terms using amplitude addition \cite{vazquez_efficient_2020}. Note that this approach requires an approximation of $D(\theta_j)$ instead of $\rho^{-1}(\theta_j)$, and likewise the consideration of probabilities instead of amplitudes in the Fourier series expansion, where the latter can be accounted for by employing squared sinusoidal functions. Using simple trigonometric identities, the first four terms of the corresponding approximation are given by
\begin{equation}\label{eq:denominator}
\begin{split}
    \frac{D(\theta_j)}{2}
    \approx
    \; &0.915 - 0.485\cos^2(2\theta_j) \\
    &+ 0.083\cos^2(4\theta_j) - 0.014\cos^2(6\theta_j).
\end{split}
\end{equation}

\noindent The factor $1/2$ must be included to ensure normalization. In principle it is possible to construct a quantum circuit that entirely computes Eq.~\eqref{eq:denominator} on the amplitude of a single qubit, and multiply it to the gearbox output according to Eq.~\eqref{eq:fixed_multi-state_measurement}, i.e.,
%
%
\begin{equation}\label{eq:circuit_splitting}
	\frac{\Omega(10)}{2}
	\approx
	\underbrace{
		\frac{1}{4} \sum_{j=0}^3 D(\theta_j)\, \rho^2(\theta_j) \sin^2\left[ \arctan[ \tan^2 \theta_j ] \right]
	}_{\text{circuit }\Delta}
\end{equation}
After multiplying the result of the measurement for $\Delta$ with the factor of $2$, indeed $\overline{\mathcal{S}^{\circ 1}}(\pmb{\theta})$ is obtained. The construction of the quantum circuit $\Delta$ including amplitude subtraction is detailed in the \SI (see Section~SI~4 and following).

The implementation of $\Delta$ theoretically requires $238$ CX-gate applications and is far beyond what we expect to be reasonably implementable on state-of-the-art NISQ hardware. Therefore, in the remaining part of this section we present a strategy for obtaining reliable, experimental results  for $\Delta$ on \IBMQ regardless. The decisive advantage from our approach for implementing the Fourier Series expansion of $D(\theta_j)$ can be demonstrated by rewriting Eq.~\eqref{eq:circuit_splitting} as
\begin{equation}\label{eq:circuit_splitting_2}
	\frac{\Omega(10)}{2}
	\approx
	\sum_{k=0}^3 (-1)^k
	\underbrace{
	\frac{1}{4} \sum_{j=0}^3 D^{2k}(\theta_j)\, \rho^2(\theta_j) \sin^2\left[ \arctan[ \tan^2 \theta_j ] \right]
    }_{\text{subcircuit } \Delta^{2k}}.
\end{equation}
where $D^{2k}(\theta_j) = a_{2k} \cos^2(2k\theta_j)$ and $a_{2k}$ stemming from Eq.~\eqref{eq:denominator}. Accordingly, instead of performing the full transformation by a single, complex circuit $\Delta$, we are able to split $\Delta$ into four subcircuits $\Delta^{2k}$, implementing the $D^{2k}(\theta_j)$ elements separately. After execution, the individual results are then post-processed as $\Delta = \Delta^0 - \Delta^2 + \Delta^4 - \Delta^6$ to eventually obtain $\overline{\mathcal{S}^{\circ 1}}(\pmb{\theta})$.
\begin{figure*}
    \centering
    \includegraphics[width=1\textwidth]{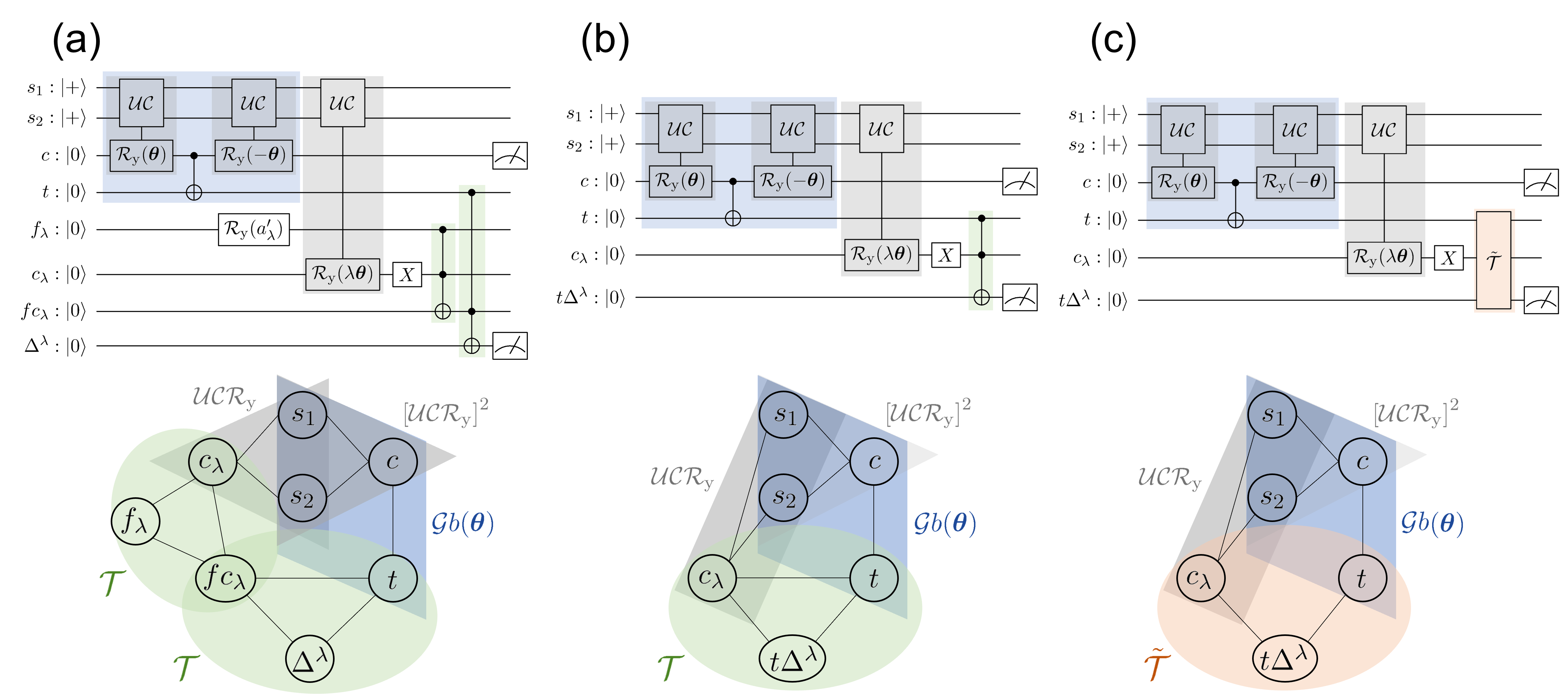}
    \caption{
    The upper row shows quantum circuits for different variants of the $\Delta^\lambda$ subcircuit family.
    (a) Full subcircuit,
    (b) subcircuit with externalized multiplication by $a_\lambda$, and
    (c) as in (b) but with the phase-equivalent version of the Toffoli gate. Generally, following the gearbox procedure, all values are encoded in the amplitude for the $\ket{1}$ state of each qubit, i.e., qubits $f_\lambda$, $c_\lambda$ and $f c_\lambda$ store the values for the coefficient $\sqrt{a_\lambda}$, the cosine term $\cos(\lambda \theta_j)$, and their product $\sqrt{a_\lambda} \cos(\lambda \theta_j)$, respectively. Qubits $\Delta^\lambda$ and $t\Delta^\lambda$ represent final circuit results, where the latter indicates that a further post-processing step is required. Coefficients $\sqrt{a_\lambda}$ are encoded using a $\mathcal{R}_\mathrm{y}(a_\lambda')$ gate with $a_\lambda' = \arcsin (\sqrt{a_\lambda})$. Topological maps in the lower row demonstrate the respectively required connectivity on a quantum computer. Lines indicate direct communication between the involved qubits. Subcircuit elements are identified using color-coding. The number of CX gates are summarized in Table~\ref{table:gate_counts}.
    }
    \label{fig:circuit_optimization}
\end{figure*}
Each subcircuit $\Delta^{2k}$ begins with the single-step gearbox as shown in Figure~\ref{fig:uniformly-controlled_implementation}b with the state register represented by $\ket{\Psi} = \ket{+}\ket{+}$. Since subcircuit $\Delta^0$ constitutes a specifically simple case, which only includes rescaling the gearbox result by $a_0$ (cf.\ Eq.~\eqref{eq:denominator} and Eq.~\eqref{eq:circuit_splitting_2}), its discussion is postponed for the moment.  
%
The $\Delta^{\lambda}, \lambda=2,4,6$ subcircuit family is more elaborate: an additional uniformly-controlled rotation is required to encode the respective term $\cos^2(\lambda \theta_j)$, which is subsequently rescaled by $a_\lambda$, and finally multiplied to the gearbox result. The corresponding quantum circuit is demonstrated in the top row of Figure~\ref{fig:circuit_optimization}a.
In the lower row, the corresponding topological requirement for a quantum computer is schematically shown using color-codes and lines for indicating direct communication between the involved qubits. Assuming all-to-all connectivity, a $\Delta^\lambda$ subcircuit requires 25 CX gates. Considering the typical heavy-hexagon architecture of \emph{IBMQ} devices, even the most efficient hardware realization found for $\Delta^\lambda$ involves 44 CX-gate applications, and thus still remains too technically demanding. The last resource available for further simplifying the $\Delta^{2k}$ subcircuits remains in externalizing the multiplication by $a_\lambda$ to a classical computer. The corresponding  quantum circuit, shown in the upper panel of Figure~\ref{fig:circuit_optimization}b, solely rescales the gearbox result with the respective $\cos^2(\lambda \theta_j)$ term, and stores the value in a qubit $t \Delta^\lambda$. The final result is then obtained post-measurement by classically computing $\Delta^\lambda = a_\lambda \cdot t \Delta^\lambda$.
This modification in turn allows further simplifying the last multiplication step implemented by the Toffoli gate $\mathcal{T}$ \cite{deBakker_reversible_1980} (green-colored box in Figure~\ref{fig:circuit_optimization}): even the most efficient implementation requires 6 CX-gate applications and all-to-all connectivity of the three involved qubits \cite{nielsen_chuang_2010, Shende_OnTC_2009}. However, a phase-equivalent version $\tilde{\mathcal{T}}$ has been reported, performing the identical transformation as $\mathcal{T}$, but additionally flagging the $\ket{101}$ state (referring to the three involved qubits) with a negative amplitude (phase shift by $\pi$) \cite{divincenzo_1994_twoBitGateDesign, Barenco_elemGates_1995}.
This version can be implemented using only $3$ CX gates, and additionally does not require communication between the two control qubits. Both hardware implementations for the Toffoli gate and the phase equivalent version can be found in the \SI (Figure~S6). Since $\mathcal{T}$ in Figure~\ref{fig:circuit_optimization}b is immediately followed by a measurement, $\tilde{\mathcal{T}}$ can be employed regardless. This is demonstrated in Figure~\ref{fig:circuit_optimization}c, and ultimately allowed us to reduce the number of CX gates to $16$ assuming all-to-all connectivity, and to $25$ considering the architecture of \IBMQ. All requirements for the quantum circuits from Figure~\ref{fig:circuit_optimization} are summarized in Table~\ref{table:gate_counts}.
\begin{table}[bt]
    \centering
    \caption{Hardware requirements for the different variants of the $\Delta^\lambda$ subcircuits shown in Figure~\ref{fig:circuit_optimization}.}
    \resizebox{1\columnwidth}{!}{
    \begin{tabular}{ c*{4}{c} }
        \multicolumn{1}{>\centering m{1.5cm}}{\multirow{2}{5.5em}{\centering $\Delta^\lambda$-subcircuit variant}} & \multicolumn{1}{>\centering m{1.5cm}}{\multirow{2}{5em}{\centering Qubits}} & \multicolumn{3}{c}{CX-gate applications} \\
        \cmidrule(lr){3-5}
        & & \multicolumn{1}{>\centering m{2cm}}{All-to-all} & \multicolumn{1}{>\centering m{2cm}}{\emph{Ehningen}} & \multicolumn{1}{>\centering m{2cm}}{Overhead} \\
        \cmidrule{1-5}
        (a) & 8 & 25 & 44 & 19 \\
        (b) & 7 & 19 & 29 & 10 \\
        (c) & 7 & 16 & 25 & 9 \\
    \end{tabular}
    }
    \label{table:gate_counts}
\end{table}
%

\noindent Note that multiplying $a_\lambda$ post-measurement can likewise be employed for $\Delta^0$, leaving the circuit at the same complexity as described in Figure~\ref{fig:state-dependent_implementation} ($9$ CX-gate applications, no overhead). In Figure~\ref{fig:multi-state_implementation_comparison}a, right panel, the analytical value $\overline{\mathcal{S}^{\circ 1}}(\pmb{\theta}) \approx 0.567$ is compared to simulated and experimental estimators for $\Omega(10)$ according to Eq.~\eqref{eq:fixed_multi-state_measurement}.
Data for $\Omega(10)$ was obtained by performing runs of the four subcircuits $\Delta^\lambda, \lambda=0,2,4,6$ in their most simplified version on an ideal, noise-free quantum computer assuming all-to-all connectivity, and on the \IBMQ device using the optimum hardware realizations. The inset indicated by the red box is again given in Figure~\ref{fig:multi-state_implementation_comparison}b. The experimental procedure is identical to the one reported for $\Omega(1 \mid 0)$, shown in the left panel.
Notably, the simulated result for $\Omega(10)$ (green diamond) is slightly larger than the analytical value (black cross). This is in accordance with stopping the series expansion of $D(\theta_j)$ after a negative term, leaving the approximation below the exact value.
Regarding the $100$ individual experimental results for $\Omega(10)$ (small gray circles), a wider scattering can be observed when compared to the experimental data for $\Omega(1\mid 0)$ (left panel); the standard deviation (blue error bar) was found about three times larger, which is expected considering the significant increase in CX-gate applications. Nevertheless, the overall estimator for $\Omega(10)$ (blue circles) is in good agreement with the analytical and simulated values, and is clearly distinguishable from the experimental data obtained for $\Omega(1\mid 0)$.

%
%

\section*{Discussion}

\noindent In this article, we have introduced an amplitude-based encoding for an approximation of the USF. This involved a quantum circuit that, based on a given input, prepares a quantum register to reflect the corresponding USF output. 
Two circuit variations were suggested receiving the input either (A) directly from a classical computer, or (B) via a quantum state when incorporated into a larger circuit. For (A), we have demonstrated different levels of approximation, and furthermore showed small circuit extension allowing to approximate other non-linear functions. For (B), we likewise presented a circuit extension for quantum-state inputs in superposition, e.g., when computing the average output of the USF. Supported by analytical and simulated data, the performance for all quantum circuits was evaluated on the IBM Quantum device in Ehningen, Germany.
Reliable experimental results were presented, obtained from quantum circuits that included up to 8 qubits, and up to 25 CX-gate applications, clearly demonstrating the applicability of our approach on state-of-the-art Noisy Intermediate-Scale Quantum devices available to date.

\section*{Methods}\label{sec:methods}

\noindent%
The python code for constructing, optimizing, and executing the quantum circuits discussed in this article was written using the Qiskit framework \cite{Qiskit}.

\subsection*{Simulation}
\noindent All simulations have been performed using the AerSimulator included in Qiskit. Throughout, an ideal, noise-free quantum computer with all-to-all connectivity was assumed. We would like to emphasize that the number of circuit executions for each input angle or state has been set to $10^5$ for maintaining comparability to the experimental data. However, sufficiently small confidence intervals were already observed for a significantly lower amount of repetitions, ranging between $2^{10}$ to $2^{14}$.

\subsection*{Hardware-Optimization Procedure}
\noindent Optimizing the quantum circuits demonstrated in this article on \IBMQ requires transpilation prior to execution. Within Qiskit terminology, transpilation can be understood as a pipeline consisting of four steps:
(1) rewriting the circuit in terms of the basis gate library of the backend,
(2) mapping the virtual qubits to the physical qubits of the device (initial layout),
(3) incorporating swap operations necessary due to the restricted connectivity of the involved physical qubits, and
(4) optimizing the employed gates.
Throughout we have chosen transpilation using the highest diligence (optimization level 3), that relies on the SWAP-based BidiREctional heuristic search algorithm (SABRE) for finding the optimum initial layout and swapping strategy \cite{li_2019_qubitMappingProblem}, and additionally perform full single- and double-qubit-gate optimizations. Due to its stochastic nature, we have repeated transpilation $50$ times, and saved the resulting transpiled circuit
with the lowest number of CX gates. Note that the corresponding virtual-to-physical qubit mapping is optimum in terms of CX-gate applications, but does not consider the quality of the involved qubits. 
Since we only use $8$ of the $27$ qubits available on \IBMQ at most, typically
this transpiled circuit can be reconstructed using different qubit subsets. The \emph{mapomatic} package \cite{nation_2022_mapomatic} allows to evaluate all of these different subsets regarding their individual error rates.
In such a way we eventually identified the optimum hardware-realization of the circuit with respect to the CX-gate applications and the quality of the involved qubits.
It is emphasized that the full procedure is only required once for a distinct circuit and does not need to be repeated when changing an input angle or state.
Post-measurement, error mitigation using the matrix-free measurement mitigation (M3) package was applied throughout \cite{nation_2021_mitigationOfError}. Every circuit was executed $10^5$ times, corresponding to the current maximum number of repetitions on the \IBMQ. It should be noted that all experiments were conducted shortly after (typically less than 30 minutes) calibration of the device. 


\bibliographystyle{naturemag}
\bibliography{lit}

\section*{Data Availability}
    \noindent The Python code used for data generation is available upon request.\\
    
\section*{Acknowledgements}
    \noindent This work was supported by the project AnQuC-3 of the Competence Center Quantum Computing Rhineland-Palatinate (Germany). We are highly indebted for being granted access to the IBM Quantum device in Ehningen, Germany.

\section*{Author Contributions}
    \noindent All authors researched, collated, and wrote this paper.
    
\section*{Competing Interests}
    \noindent The authors declare no competing interests.
    
\section*{Additional Information}
    \noindent This version is supported by supplementary material.


\end{document}


\title{Supplementary Information:\\An Amplitude-Based Implementation of the Unit Step Function on a Quantum Computer}

\author{Jonas Koppe}
\email[Correspondence: Jonas Koppe ]{(jonas.koppe@itwm.fraunhofer.de)}
\affiliation{Department of Financial Mathematics, Fraunhofer ITWM}

\author{Mark-Oliver Wolf}
\affiliation{Department of Financial Mathematics, Fraunhofer ITWM}

\maketitle

\patchcmd{\section}
    {\centering}{\raggedright}{}{}
 \patchcmd{\subsection}
    {\centering}{\raggedright}{}{}

\documentclass[../main.tex]{subfiles}
\graphicspath{{images/}{../images/}{../images/SI images}}%
\begin{document}

\patchcmd{\section}
  {\centering}
  {\raggedright}
  {}
  {}

\tableofcontents

\newpage
\documentclass[../main.tex]{subfiles}
\graphicspath{{images/}{../images/}{../images/SI images}}%
\begin{document}

\section{Success Probability for $d$-Step Gearbox Circuits}\label{SI:success_prob}

\noindent Extending the discussion given in the main text, the transformation performed by a $d$-step gearbox circuit can be expressed as
\begin{equation}\label{eq:si_gearbox_nested}
	\Gb^{d}(\theta) \ket{0^{\otimes 2^d-1}}_c \ket{0}_t =
	\rho_{d} (\theta) \ket{0^{\otimes 2^d-1}}_c \; e^{ -i \arctan \left[  \tan^{2^d} \theta \right] X } \ket{0}_t
	- \sqrt{ 1 - \rho^2(\theta) } \ket{\emptyset^{\otimes 2^d-1}}_c \; e^{ -i \frac{\pi}{4} X } \ket{0}_t,
\end{equation}
where $\rho^2_{d} (\theta)$ represents the corresponding $d$-step success probability. Even without knowing the exact expression for $\rho^2_{d} (\theta)$,
we can derive the result for post-measurement data evaluation,
\begin{equation}\label{eq:si_cond_prob}
	\begin{split}
		\Omega ( 1 \mid 0^{\otimes 2^d-1} )
		&=
		\Prob \left( \ket{1}_t \mid \ket{0^{\otimes 2^d-1}}_c \right)
		=
		\frac{
			\left| \left( \bra{1}_t \bra{0^{\otimes 2^d-1}}_c \right) \ket{\tau_{\Gb}} \right|^2
		}{
			\left| \left( \bra{0}_t \bra{0^{\otimes 2^d-1}}_c \right) \ket{\tau_{\Gb}} \right|^2
			+
			\left| \left( \bra{1}_t \bra{0^{\otimes 2^d-1}}_c \right) \ket{\tau_{\Gb}} \right|^2
		} \\
		&=
		\frac{
			\rho_{d}^2(\theta) \sin^2 \left[ \arctan [ \tan^{2^d} \theta ] \right]
		}{
			\rho_{d}^2(\theta) \cos^2 \left[ \arctan [ \tan^{2^d} \theta ] \right]
			+
			\rho_{d}^2(\theta) \sin^2 \left[ \arctan [ \tan^{2^d} \theta ] \right]
		} \\
		&= 
		\sin^2 \left[ \arctan [ \tan^{2^d} \theta ] \right]
		= \mathcal{S}^{\circ d}(\theta),
	\end{split}
\end{equation}
in accordance with the simulated and experimental data presented in the main text. The transformation and subsequent data evaluation may likewise be expressed avoiding the decomposition with respect to the control register, i.e., into 
its all-zero state $\ket{0^{\otimes 2^d-1}}_c$ and all other states $\ket{\emptyset^{\otimes 2^d-1}}_c$ as demonstrated in Eq.~\eqref{eq:si_gearbox_nested}: the final state of all qubits involved in a $d$-step gearbox can be described by $2^{d+1}$-dimensional vector with the elements zero and $2^{d}$ given by
\begin{equation}\label{eq:si_gb_vector_alt}
	\Gb^{d}(\theta) \ket{0^{\otimes 2^d-1}}_c \ket{0}_t=
	\begin{pmatrix}
		\cos^{2^{d}}(\theta)\\
		\vdots\\
		\sin^{2^{d}}(\theta)\\
		\vdots
	\end{pmatrix}
	=\ket{\tau_{\Gb}},
\end{equation}
where using Qiskit \cite{Qiskit} ordering these elements represent the amplitudes for the states $\ket{0}_t \ket{0^{\otimes 2^d-1}}_c$ and $\ket{1}_t \ket{0^{\otimes 2^d-1}}_c$, respectively.
In contrast to Eq.~\eqref{eq:si_gearbox_nested}, this expression can simply be derived from linear algebra. For instance, the transformation for the single-step gearbox with $d=1$ is given by
\begin{equation}\label{eq:si_sgb_exc}
	\begin{pmatrix}
		\cos^{2}(\theta) & \sin(\theta)\cos(\theta) & \sin^{2}(\theta) & -\sin(\theta)\cos(\theta)\\
		\sin(\theta)\cos(\theta) & \sin^{2}(\theta) & -\sin(\theta)\cos(\theta) & \cos^{2}(\theta)\\
		\sin^{2}(\theta) & -\sin(\theta)\cos(\theta) & \cos^{2}(\theta) & \sin(\theta)\cos(\theta)\\
		-\sin(\theta)\cos(\theta) & \cos^{2}(\theta) & \sin(\theta)\cos(\theta) & \sin^{2}(\theta)\\
	\end{pmatrix}
	\begin{pmatrix}
		1\\
		0\\
		0\\
		0\\
	\end{pmatrix}
	=	
	\begin{pmatrix}
		\cos^{2}(\theta)\\
		\sin(\theta)\cos(\theta)\\
		\sin^{2}(\theta)\\
		-\sin(\theta)\cos(\theta)
	\end{pmatrix}
\end{equation}
Using the alternative expression for $\ket{\tau_{\Gb}}$ as given in Eq.~\eqref{eq:si_gb_vector_alt} for the data evaluation according to the first line of Eq.~\eqref{eq:si_cond_prob}, we obtain
\begin{equation}\label{eq:si_cond_prob_alter}
	\begin{split}
		\Omega ( 1 \mid 0^{\otimes 2^d-1} )
		&=
		\frac{
			\sin^{2^{d+1}}(\theta)
		}{
			\cos^{2^{d+1}}(\theta) + \sin^{2^{d+1}}(\theta)
		}
		= \mathcal{S}^{\circ d}(\theta).
	\end{split}
\end{equation}
Comparing the terms for $\left| \left( \bra{1}_t \bra{0^{\otimes 2^d-1}}_c \right) \ket{\tau_{\Gb}} \right|^2$ derived from Eqs.~\eqref{eq:si_cond_prob} and \eqref{eq:si_cond_prob_alter}, we obtain the $d$-step success probability
\begin{equation}\label{eq:si_gb_sprob_d_step}
	\rho^2_{d} (\theta) = \sin^{2^{d+1}}(\theta) + \cos^{2^{d+1}}(\theta).
\end{equation}
Consider the objective to calculate average USF outputs using $d$-step gearbox circuits with quantum-state inputs. Note that the inverse $D_d(\theta)$ of Eq.~\eqref{eq:si_gb_sprob_d_step} has to be approximated using Fourier expansion. The resulting terms can then be implemented as demonstrated in the last part of the main text. Using $\cos(2x)=2\cos^2(x)-1$ again, the first five terms from the expansion for the double-step gearbox with $d=2$ are given by
\begin{equation}
	\frac{D_2(\theta)}{8}
	\approx
	0.598 - 0.7\cos^2(2\theta)
	+ 0.314\cos^2(4\theta) - 0.14\cos^2(6\theta)+0.062\cos^2(8\theta).
\end{equation}
Here, the factor of $1/8$ must be included to ensure normalization. Generally, the normalization constant $N$ can be calculated according to $N=2^{-2^d+1}$.

\end{document}
\newpage

\documentclass[../main.tex]{subfiles}
\graphicspath{{images/}{../images/}{../images/SI images}}%
\begin{document}

\section{Optimum Quantum Circuits on \IBMQ}\label{SI:exp_circs}

\noindent \textbf{The \IBMQ Device}

\begin{figure}[h!]
	\begin{center}
		\includegraphics[width=0.5\textwidth]{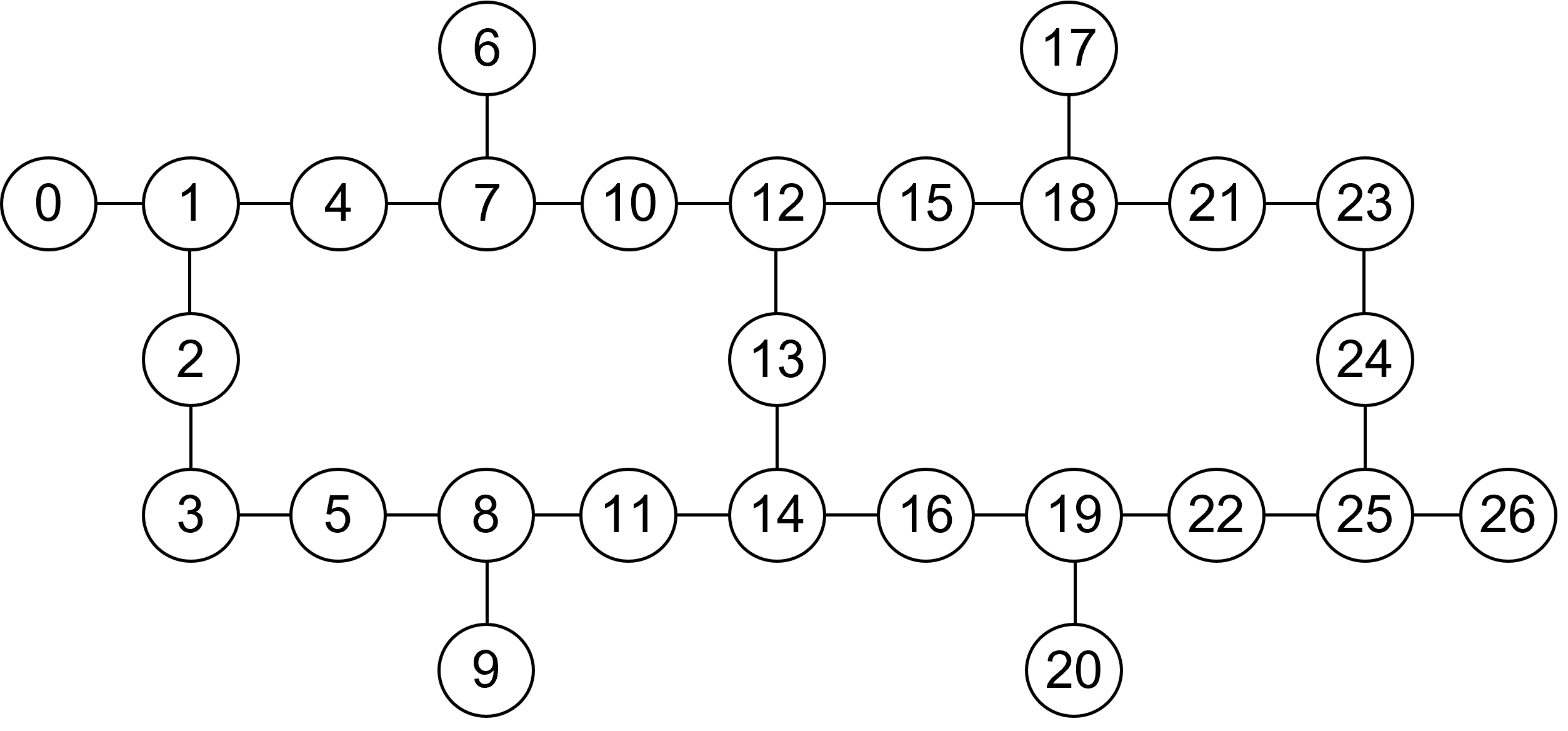}
		\caption{Heavy-hexagon architecture of the \IBMQ device.}
	    \label{fig:SI_ehningen}
	\end{center}
\end{figure}

\noindent At the time of experimental implementation, the \IBMQ device was equipped with Falcon r5.11 processor type. The average CX-gate error was determined to be about $10^{-2}$. Average relaxation times were found about $150\,\mu$s ($\mathrm{T}_1\sim160\,\mu$s and $\mathrm{T}_2\sim130\,\mu$s). Figure~\ref{fig:SI_ehningen} demonstrate the typical heavy-hexagon architecture of the $27$ qubits. The qubit numbering is used to identify physical qubits in the virtual-to-physical qubit mapping shown in the upcoming figures of the section.

\noindent \textbf{Double-Step Gearbox with rescaled first plateau}

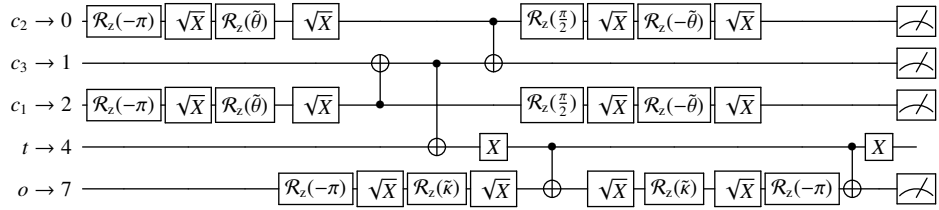
\begin{figure}[h!]
	\begin{center}
		\scalebox{0.9}{
			\Qcircuit @C=0.2em @R=0.3em @!R {					
				\lstick{c_2 \rightarrow 0} & \gate{\mathcal{R}_{\mathrm{z}}(-\pi)} & \gate{\sqrt{X}} & \gate{\mathcal{R}_{\mathrm{z}}(\tilde{\theta})} & \gate{\sqrt{X}} & \qw & \qw & \ctrl{1} & \gate{\mathcal{R}_{\mathrm{z}}(\frac{\pi}{2})} & \gate{\sqrt{X}} & \gate{\mathcal{R}_{\mathrm{z}}(-\tilde{\theta})} & \gate{\sqrt{X}} & \qw & \qw & \qw & \meter\\	
				\lstick{c_3 \rightarrow 1} & \qw & \qw & \qw & \qw & \targ & \ctrl{2} & \targ & \qw & \qw & \qw & \qw & \qw & \qw & \qw & \meter \\
				\lstick{c_1 \rightarrow 2} & \gate{\mathcal{R}_{\mathrm{z}}(-\pi)} & \gate{\sqrt{X}} & \gate{\mathcal{R}_{\mathrm{z}}(\tilde{\theta})} & \gate{\sqrt{X}} & \ctrl{-1} & \qw & \qw & \gate{\mathcal{R}_{\mathrm{z}}(\frac{\pi}{2})} & \gate{\sqrt{X}} & \gate{\mathcal{R}_{\mathrm{z}}(-\tilde{\theta})} & \gate{\sqrt{X}} & \qw & \qw & \qw & \meter\\
				\lstick{t \rightarrow 4} & \qw & \qw & \qw & \qw & \qw & \targ & \gate{X} & \ctrl{1} & \qw & \qw & \qw & \qw & \ctrl{1} & \gate{X} & \qw \\
				\lstick{o \rightarrow 7} & \qw & \qw & \qw & \gate{\mathcal{R}_{\mathrm{z}}(-\pi)} & \gate{\sqrt{X}} & \gate{\mathcal{R}_{\mathrm{z}}(\tilde{\kappa})} & \gate{\sqrt{X}} & \targ & \gate{\sqrt{X}} & \gate{\mathcal{R}_{\mathrm{z}}(\tilde{\kappa})} & \gate{\sqrt{X}} & \gate{\mathcal{R}_{\mathrm{z}}(-\pi)} & \targ & \qw & \meter\\	
			}
		}
	\end{center}
	\caption{Optimum quantum circuit for \IBMQ, corresponding to circuit shown in Figure~5b 
	in the main text. Here we used $\tilde{\theta}=\pi-\theta$, and $\tilde{\kappa}=\pi-\kappa$ with $\kappa=\pi/4$. Moreover, the first CX gate involving the target qubit $t$ and the output qubit $o$ can be avoided by actually using $\tilde{\kappa}$ in the $\mathcal{R}_{\mathrm{z}}(\kappa)$ gate instead of $\kappa$, and evaluating the measurement result according to $\Omega ( 0 \mid 0^{\otimes 3} )$, following the definition given in Eq.~\eqref{eq:si_cond_prob}, i.e., replacing $\left| \left( \bra{1}_t \bra{0^{\otimes 3}}_c \right) \ket{\tau_{\Gb}} \right|^2$ by $\left| \left( \bra{0}_t \bra{0^{\otimes 3}}_c \right) \ket{\tau_{\Gb}} \right|^2$ in the numerator. The first column indicates the virtual-to-physical qubit mapping.}
	\label{fig::exp_circ_dbg_rescale}
\end{figure}

\noindent \textbf{ReLU Using Double-Step Gearbox}

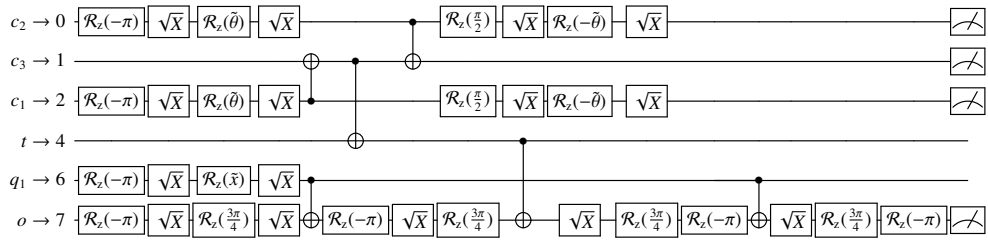
\begin{figure}[h!]
	\begin{center}
		\scalebox{0.8}{
			\Qcircuit @C=0.2em @R=0.3em @!R {					
				\lstick{c_2 \rightarrow 0} & \gate{\mathcal{R}_{\mathrm{z}}(-\pi)} & \gate{\sqrt{X}} & \gate{\mathcal{R}_{\mathrm{z}}(\tilde{\theta})} & \gate{\sqrt{X}} & \qw & \qw & \ctrl{1} & \gate{\mathcal{R}_{\mathrm{z}}(\frac{\pi}{2})} & \gate{\sqrt{X}} & \gate{\mathcal{R}_{\mathrm{z}}(-\tilde{\theta})} & \gate{\sqrt{X}} & \qw & \qw & \qw & \qw & \qw & \meter\\	
				\lstick{c_3 \rightarrow 1} & \qw & \qw & \qw & \qw & \targ & \ctrl{2} & \targ & \qw & \qw & \qw & \qw & \qw & \qw & \qw & \qw & \qw & \meter \\
				\lstick{c_1 \rightarrow 2} & \gate{\mathcal{R}_{\mathrm{z}}(-\pi)} & \gate{\sqrt{X}} & \gate{\mathcal{R}_{\mathrm{z}}(\tilde{\theta})} & \gate{\sqrt{X}} & \ctrl{-1} & \qw & \qw & \gate{\mathcal{R}_{\mathrm{z}}(\frac{\pi}{2})} & \gate{\sqrt{X}} & \gate{\mathcal{R}_{\mathrm{z}}(-\tilde{\theta})} & \gate{\sqrt{X}} & \qw & \qw & \qw & \qw & \qw & \meter\\
				\lstick{t \rightarrow 4} & \qw & \qw & \qw & \qw & \qw & \targ & \qw & \qw & \ctrl{2} & \qw & \qw & \qw & \qw & \qw & \qw & \qw & \qw \\
				\lstick{q_1 \rightarrow 6} & \gate{\mathcal{R}_{\mathrm{z}}(-\pi)} & \gate{\sqrt{X}} & \gate{\mathcal{R}_{\mathrm{z}}(\tilde{x})} & \gate{\sqrt{X}} & \ctrl{1} & \qw & \qw & \qw & \qw & \qw & \qw & \qw & \ctrl{1}\qw & \qw & \qw & \qw & \qw\\	
				\lstick{o \rightarrow 7} & \gate{\mathcal{R}_{\mathrm{z}}(-\pi)} & \gate{\sqrt{X}} & \gate{\mathcal{R}_{\mathrm{z}}(\frac{3\pi}{4})} & \gate{\sqrt{X}} & \targ & \gate{\mathcal{R}_{\mathrm{z}}(-\pi)} & \gate{\sqrt{X}} & \gate{\mathcal{R}_{\mathrm{z}}(\frac{3\pi}{4})} & \targ & \gate{\sqrt{X}} & \gate{\mathcal{R}_{\mathrm{z}}(\frac{3\pi}{4})} & \gate{\mathcal{R}_{\mathrm{z}}(-\pi)} & \targ & \gate{\sqrt{X}} & \gate{\mathcal{R}_{\mathrm{z}}(\frac{3\pi}{4})} & \gate{\mathcal{R}_{\mathrm{z}}(-\pi)} & \meter\\	
			}
		}
	\end{center}
	\caption{Optimum quantum circuit for \IBMQ implementing the ReLU function, corresponding to the circuit shown in Figure~5d 
	in the main text. Here we used $\tilde{x}=\pi-x$ with the definition for $x$ as given in the main text. The first column indicates the virtual-to-physical qubit mapping.}
	\label{fig::exp_circ_dbg_relu}
\end{figure}

\newpage

\begin{sidewaysfigure}[tb]
	\centering
	\textbf{(a) Double-Step Gearbox}
	\newline
	\scalebox{.8}{
		\Qcircuit @C=0.5em @R=0.5em @!R {
			\lstick{c_1 \rightarrow 15} & \gate{\mathcal{R}_{\mathrm{z}}(-\pi)} & \gate{\sqrt{X}} & \gate{\mathcal{R}_{\mathrm{z}}(\tilde{\theta})} & \gate{\sqrt{X}} & \ctrl{2} & \qw & \qw & \gate{\mathcal{R}_{\mathrm{z}}(\frac{\pi}{2})} & \gate{\sqrt{X}} & \gate{\mathcal{R}_{\mathrm{z}}(-\tilde{\theta})} & \gate{\sqrt{X}} & \meter\\ 
			\lstick{c_2 \rightarrow 17} & \gate{\mathcal{R}_{\mathrm{z}}(-\pi)} & \gate{\sqrt{X}} & \gate{\mathcal{R}_{\mathrm{z}}(\tilde{\theta})} & \gate{\sqrt{X}} & \qw & \qw & \ctrl{1} & \gate{\mathcal{R}_{\mathrm{z}}(\frac{\pi}{2})} & \gate{\sqrt{X}} & \gate{\mathcal{R}_{\mathrm{z}}(-\tilde{\theta})} & \gate{\sqrt{X}} & \meter\\
			\lstick{c_3 \rightarrow 18} & \qw & \qw & \qw & \qw & \targ & \ctrl{1} & \targ & \qw & \qw & \qw & \qw & \meter \\
			\lstick{t \rightarrow 21} & \qw & \qw & \qw & \qw & \qw & \targ & \qw & \qw & \qw & \qw & \qw & \meter
		}
	}\newline
	\newline%
	\textbf{(b) Triple-Step Gearbox}
	\newline
	\scalebox{.73}{ 
		\Qcircuit @C=0.2em @R=0.6em @!R {
			\lstick{c_1 \rightarrow 0} & \gate{\mathcal{R}_{\mathrm{z}}(-\pi)} & \gate{\sqrt{X}} & \gate{\mathcal{R}_{\mathrm{z}}(\tilde{\theta})} & \gate{\sqrt{X}} & \ctrl{1} & \qw & \qw & \gate{\mathcal{R}_{\mathrm{z}}(\frac{\pi}{2})} & \gate{\sqrt{X}} & \gate{\mathcal{R}_{\mathrm{z}}(-\tilde{\theta})} & \gate{\sqrt{X}} & \qw & \qw & \qw & \qw & \qw & \qw & \qw & \qw & \qw & \qw & \qw & \qw & \qw & \qw & \meter\\ 
			\lstick{c_5 \rightarrow 1} & \qw & \qw & \qw & \qw & \targ & \ctrl{2} & \gate{\mathcal{R}_{\mathrm{z}}(-\frac{\pi}{2})} & \targ & \qw & \qw & \qw & \qw & \qw & \qw & \qw & \qw & \qw & \qw & \qw & \qw & \qw & \qw & \qw & \qw & \qw & \meter\\
			\lstick{c_2 \rightarrow 2} & \gate{\mathcal{R}_{\mathrm{z}}(-\pi)} & \gate{\sqrt{X}} & \gate{\mathcal{R}_{\mathrm{z}}(\tilde{\theta})} & \gate{\sqrt{X}} & \qw & \qw & \qw & \ctrl{-1} & \gate{\mathcal{R}_{\mathrm{z}}(-\frac{\pi}{2})} & \gate{\sqrt{X}} & \gate{\mathcal{R}_{\mathrm{z}}(\tilde{\theta})} & \gate{\sqrt{X}} \qw & \qw & \qw &  \qw &  \qw & \qw & \qw & \qw & \qw & \qw & \qw & \qw & \qw & \qw & \meter\\
			\lstick{c_7 \rightarrow 4} & \qw & \qw & \qw & \qw & \qw & \targ & \ctrl{2} & \targ & \ctrl{2} & \qw & \qw & \qw & \qw & \qw & \qw & \qw & \qw & \qw & \qw & \qw & \ctrl{2}  & \gate{\mathcal{R}_{\mathrm{z}}(-\frac{\pi}{2})} & \gate{\sqrt{X}} & \gate{\mathcal{R}_{\mathrm{z}}(-\tilde{\theta})} & \gate{\sqrt{X}} & \meter\\
			\lstick{t \rightarrow 6} & \qw & \qw & \qw & \qw & \qw & \qw & \qw & \qw & \qw & \targ & \qw & \qw & \qw & \qw & \qw & \qw  & \qw & \qw & \qw & \qw & \qw & \qw & \qw & \qw & \qw & \meter\\
			\lstick{c_4 \rightarrow 7} & \gate{\mathcal{R}_{\mathrm{z}}(-\pi)} & \gate{\sqrt{X}} & \gate{\mathcal{R}_{\mathrm{z}}(\tilde{\theta})} & \gate{\sqrt{X}} & \qw & \qw & \targ & \ctrl{-2} & \targ & \ctrl{-1} & \gate{\mathcal{R}_{\mathrm{z}}(-\pi)} & \gate{X} & \ctrl{1} & \gate{\sqrt{X}} & \gate{\mathcal{R}_{\mathrm{z}}(-\frac{\pi}{2})} & \gate{\sqrt{X}} & \ctrl{1} & \gate{\mathcal{R}_{\mathrm{z}}(-\frac{\pi}{2})} & \gate{\sqrt{X}} & \gate{\mathcal{R}_{\mathrm{z}}(-1.61)} & \targ & \qw &\qw &\qw &\qw  & \meter\\
			\lstick{c_6 \rightarrow 10} & \qw & \qw & \qw &\qw & \targ & \gate{\mathcal{R}_{\mathrm{z}}(-1.54)} & \gate{\sqrt{X}} & \qw & \qw & \qw & \qw & \qw & \targ & \gate{\mathcal{R}_{\mathrm{z}}(\frac{\pi}{2})} & \qw & \qw & \targ & \gate{\mathcal{R}_{\mathrm{z}}(-\frac{\pi}{2})} & \gate{\sqrt{X}} & \qw & \qw & \qw &\qw &\qw &\qw & \meter\\
			\lstick{c_3 \rightarrow 12} & \gate{\mathcal{R}_{\mathrm{z}}(-\pi)} & \gate{\sqrt{X}} & \gate{\mathcal{R}_{\mathrm{z}}(\tilde{\theta})} & \gate{\sqrt{X}} & \ctrl{-1} & \gate{\mathcal{R}_{\mathrm{z}}(\frac{\pi}{2})} & \gate{\sqrt{X}} & \gate{\mathcal{R}_{\mathrm{z}}(-\tilde{\theta})} & \gate{\sqrt{X}} & \qw & \qw & \qw & \qw & \qw & \qw & \qw  & \qw & \qw & \qw & \qw & \qw & \qw & \qw & \qw & \qw & \meter\\
		}
	}%
	\caption{Optimum quantum circuits for \IBMQ implementing (a) the double-step, and (b) the triple-step gearbox. The corresponding theoretical quantum circuits are shown in Figure~3b 
	and c in the main text. Here we used $\tilde{\theta}=\pi-\theta$. The first columns indicate the respective virtual-to-physical qubit mapping.}  
	\label{fig::exp_circ_dbg_tgb}
\end{sidewaysfigure}

\newpage

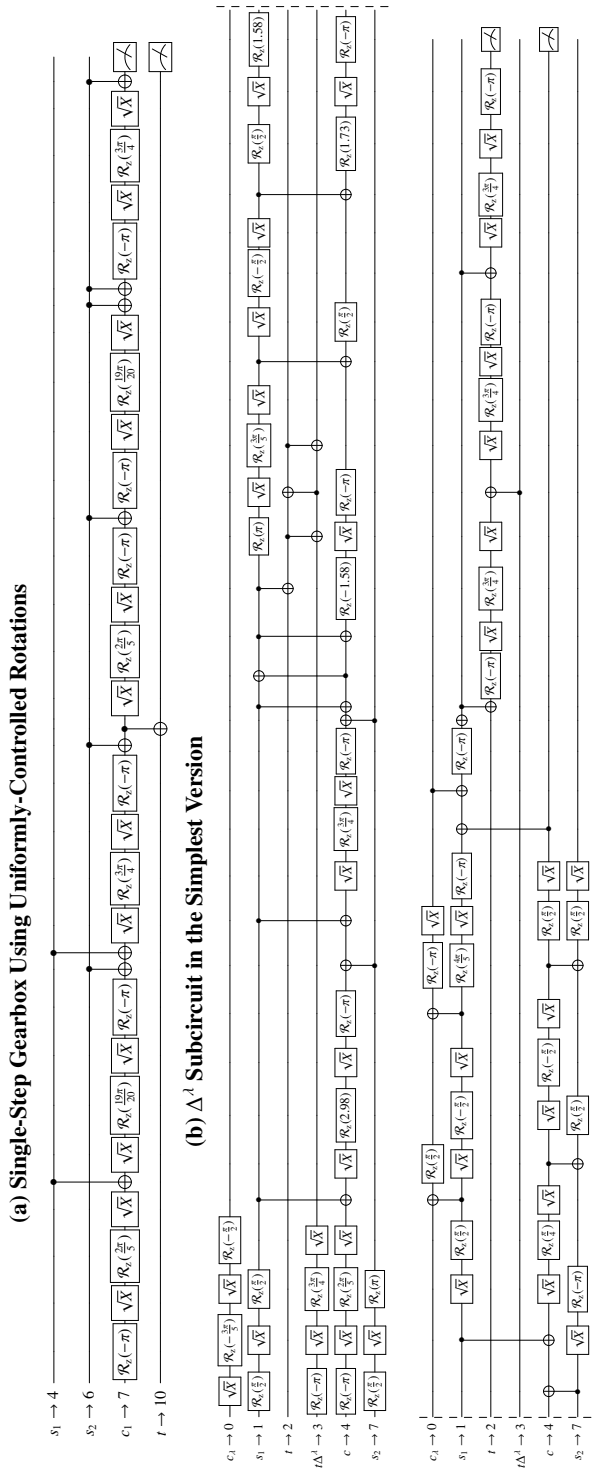
\begin{sidewaysfigure}[tb]
	\centering
	\textbf{(a) Single-Step Gearbox Using Uniformly-Controlled Rotations}
	\newline
	\scalebox{.68}{
	    \hspace{2cm} 
		\Qcircuit @C=0.2em @R=0.4em @!R {
			\lstick{s_1 \rightarrow 4} & \qw & \qw & \qw & \qw & \ctrl{2} & \qw & \qw & \qw & \qw & \qw & \ctrl{2} & \qw & \qw & \qw & \qw & \qw & \qw & \qw & \qw & \qw & \qw & \qw & \qw & \qw & \qw & \qw & \qw & \qw & \qw & \qw & \qw & \qw & \qw & \qw & \qw \\
			\lstick{s_2 \rightarrow 6} & \qw & \qw & \qw & \qw & \qw & \qw & \qw & \qw & \qw & \ctrl{1} & \qw & \qw & \qw & \qw & \qw & \ctrl{1} & \qw & \qw & \qw & \qw & \qw & \qw & \ctrl{1} & \qw & \qw & \qw & \qw & \ctrl{1} & \ctrl{1} & \qw & \qw & \qw & \qw & \ctrl{1} & \qw \\
			\lstick{c_1 \rightarrow 7} & \gate{\mathcal{R}_{\mathrm{z}}(-\pi)} & \gate{\sqrt{X}} & \gate{\mathcal{R}_{\mathrm{z}}(\frac{2\pi}{5})} & \gate{\sqrt{X}} & \targ & \gate{\sqrt{X}} & \gate{\mathcal{R}_{\mathrm{z}}(\frac{19\pi}{20})} & \gate{\sqrt{X}} & \gate{\mathcal{R}_{\mathrm{z}}(-\pi)} & \targ & \targ & \gate{\sqrt{X}} & \gate{\mathcal{R}_{\mathrm{z}}(\frac{3\pi}{4})} & \gate{\sqrt{X}} & \gate{\mathcal{R}_{\mathrm{z}}(-\pi)} & \targ & \ctrl{1} & \qw & \gate{\sqrt{X}} & \gate{\mathcal{R}_{\mathrm{z}}(\frac{2\pi}{5})} & \gate{\sqrt{X}} & \gate{\mathcal{R}_{\mathrm{z}}(-\pi)} & \targ & \gate{\mathcal{R}_{\mathrm{z}}(-\pi)} & \gate{\sqrt{X}} & \gate{\mathcal{R}_{\mathrm{z}}(\frac{19\pi}{20})} & \gate{\sqrt{X}} & \targ & \targ & \gate{\mathcal{R}_{\mathrm{z}}(-\pi)} & \gate{\sqrt{X}} & \gate{\mathcal{R}_{\mathrm{z}}(\frac{3\pi}{4})} & \gate{\sqrt{X}} & \targ & \meter \\
			\lstick{t \rightarrow 10} & \qw & \qw & \qw & \qw & \qw & \qw & \qw & \qw & \qw & \qw & \qw & \qw & \qw & \qw & \qw & \qw & \targ & \qw & \qw & \qw & \qw & \qw & \qw & \qw & \qw & \qw & \qw & \qw & \qw & \qw & \qw & \qw & \qw & \qw & \meter
		}
	}\newline
	\newline
	\textbf{(b) $\Delta^\lambda$ Subcircuit in the Simplest Version}
	\newline
	\scalebox{.555}{
		\Qcircuit @C=0.2em @R=0.4em @!R {
			\lstick{c_{\lambda} \rightarrow 0} & \gate{\sqrt{X}} & \gate{\mathcal{R}_{\mathrm{z}}(-\frac{3\pi}{5})} & \gate{\sqrt{X}} & \gate{\mathcal{R}_{\mathrm{z}}(-\frac{\pi}{2})} \barrier[90.4em]{5} & \qw & \qw & \qw & \qw & \qw & \qw & \qw & \qw & \qw & \qw & \qw & \qw & \qw & \qw & \qw & \qw & \qw & \qw & \qw & \qw & \qw & \qw & \qw & \qw & \qw & \qw & \qw & \qw & \qw  \\
			\lstick{s_1 \rightarrow 1} & \gate{\mathcal{R}_{\mathrm{z}}(\frac{\pi}{2})} & \gate{\sqrt{X}} & \gate{\mathcal{R}_{\mathrm{z}}(\frac{\pi}{2})} & \qw & \ctrl{3} & \qw & \qw & \qw & \qw & \qw & \ctrl{3} & \qw & \qw & \qw & \qw & \qw & \ctrl{3} & \targ & \ctrl{3} & \ctrl{1} & \gate{\mathcal{R}_{\mathrm{z}}(\pi)} & \gate{\sqrt{X}} & \gate{\mathcal{R}_{\mathrm{z}}(\frac{3\pi}{5})} & \gate{\sqrt{X}} & \ctrl{3} & \gate{\sqrt{X}} & \gate{\mathcal{R}_{\mathrm{z}}(-\frac{\pi}{2})} & \gate{\sqrt{X}} & \ctrl{3} & \gate{\mathcal{R}_{\mathrm{z}}(\frac{\pi}{2})} & \gate{\sqrt{X}} & \gate{\mathcal{R}_{\mathrm{z}}(1.58)} & \qw\\
			\lstick{t \rightarrow 2} & \qw & \qw & \qw & \qw & \qw & \qw & \qw & \qw & \qw & \qw & \qw & \qw & \qw & \qw & \qw & \qw & \qw & \qw & \qw & \targ & \ctrl{1} & \targ & \ctrl{1} & \qw & \qw & \qw & \qw & \qw & \qw & \qw & \qw & \qw & \qw\\
			\lstick{t\Delta^{\lambda} \rightarrow 3} & \gate{\mathcal{R}_{\mathrm{z}}(-\pi)} & \gate{\sqrt{X}} & \gate{\mathcal{R}_{\mathrm{z}}(\frac{3\pi}{4})} & \gate{\sqrt{X}} & \qw & \qw & \qw & \qw & \qw & \qw & \qw & \qw & \qw & \qw & \qw & \qw & \qw & \qw & \qw & \qw & \targ & \ctrl{-1} & \targ & \qw & \qw & \qw & \qw & \qw & \qw & \qw & \qw & \qw & \qw\\
			\lstick{c \rightarrow 4} & \gate{\mathcal{R}_{\mathrm{z}}(-\pi)} & \gate{\sqrt{X}} & \gate{\mathcal{R}_{\mathrm{z}}(\frac{2\pi}{5})} & \gate{\sqrt{X}} & \targ & \gate{\sqrt{X}} & \gate{\mathcal{R}_{\mathrm{z}}(2.98)} & \gate{\sqrt{X}} & \gate{\mathcal{R}_{\mathrm{z}}(-\pi)} & \targ & \targ & \gate{\sqrt{X}} & \gate{\mathcal{R}_{\mathrm{z}}(\frac{3\pi}{4})} & \gate{\sqrt{X}} & \gate{\mathcal{R}_{\mathrm{z}}(-\pi)} & \targ & \targ & \ctrl{-3} & \targ & \gate{\mathcal{R}_{\mathrm{z}}(-1.58)} & \gate{\sqrt{X}} & \gate{\mathcal{R}_{\mathrm{z}}(-\pi)} & \qw & \qw & \targ & \gate{\mathcal{R}_{\mathrm{z}}(\frac{\pi}{2})} & \qw & \qw & \targ & \gate{\mathcal{R}_{\mathrm{z}}(1.73)} & \gate{\sqrt{X}} & \gate{\mathcal{R}_{\mathrm{z}}(-\pi)} & \qw\\
			\lstick{s_2 \rightarrow 7} & \gate{\mathcal{R}_{\mathrm{z}}(\frac{\pi}{2})} & \gate{\sqrt{X}} & \gate{\mathcal{R}_{\mathrm{z}}(\pi)} & \qw & \qw & \qw & \qw & \qw & \qw & \ctrl{-1} & \qw & \qw & \qw & \qw & \qw & \ctrl{-1} & \qw & \qw & \qw & \qw & \qw & \qw & \qw & \qw & \qw & \qw & \qw & \qw & \qw & \qw & \qw & \qw & \qw\\
			 &  &  &  &  &  &  &  &  &  &  &  &  &  &  &  &  &  &  &  &  &  &  &  &  &  &  &  &  &  &  &  &  \\
			\lstick{c_{\lambda} \rightarrow 0} & \qw & \qw & \qw & \qw & \targ \barrier[-19.2em]{5} & \gate{\mathcal{R}_{\mathrm{z}}(\frac{\pi}{2})} & \qw & \qw & \targ & \gate{\mathcal{R}_{\mathrm{z}}(-\pi)} & \gate{\sqrt{X}} & \qw & \qw & \ctrl{1} & \qw & \qw & \qw & \qw & \qw & \qw & \qw & \qw & \qw & \qw & \qw & \qw & \qw & \qw & \qw & \qw & \qw & \qw\\
			\lstick{s_1 \rightarrow 1} & \qw & \ctrl{3} & \gate{\sqrt{X}} & \gate{\mathcal{R}_{\mathrm{z}}(\frac{\pi}{2})} & \ctrl{-1} & \gate{\sqrt{X}} & \gate{\mathcal{R}_{\mathrm{z}}(-\frac{\pi}{2})} & \gate{\sqrt{X}} & \ctrl{-1} & \gate{\mathcal{R}_{\mathrm{z}}(\frac{4\pi}{5})} & \gate{\sqrt{X}} & \gate{\mathcal{R}_{\mathrm{z}}(-\pi)} & \targ & \targ & \gate{\mathcal{R}_{\mathrm{z}}(-\pi)} & \targ & \ctrl{1} & \qw & \qw & \qw & \qw & \qw & \qw & \qw & \qw & \qw & \ctrl{1} & \qw & \qw & \qw & \qw & \qw\\
			\lstick{t \rightarrow 2} & \qw & \qw & \qw & \qw & \qw & \qw & \qw & \qw & \qw & \qw & \qw & \qw & \qw & \qw & \qw & \qw & \targ & \gate{\mathcal{R}_{\mathrm{z}}(-\pi)} & \gate{\sqrt{X}} & \gate{\mathcal{R}_{\mathrm{z}}(\frac{3\pi}{4})} & \gate{\sqrt{X}} & \targ & \gate{\sqrt{X}} & \gate{\mathcal{R}_{\mathrm{z}}(\frac{3\pi}{4})} & \gate{\sqrt{X}} & \gate{\mathcal{R}_{\mathrm{z}}(-\pi)} & \targ & \gate{\sqrt{X}} & \gate{\mathcal{R}_{\mathrm{z}}(\frac{3\pi}{4})} & \gate{\sqrt{X}} & \gate{\mathcal{R}_{\mathrm{z}}(-\pi)} & \meter \\
			\lstick{t\Delta^{\lambda} \rightarrow 3} & \qw & \qw & \qw & \qw & \qw & \qw & \qw & \qw & \qw & \qw & \qw & \qw & \qw & \qw & \qw & \qw & \qw & \qw & \qw & \qw & \qw & \ctrl{-1} & \qw & \qw & \qw & \qw & \qw & \qw & \qw & \qw & \qw & \qw\\
			\lstick{c \rightarrow 4} & \targ & \targ & \gate{\sqrt{X}} & \gate{\mathcal{R}_{\mathrm{z}}(\frac{\pi}{4})} & \gate{\sqrt{X}} & \ctrl{1} & \gate{\sqrt{X}} & \gate{\mathcal{R}_{\mathrm{z}}(-\frac{\pi}{2})} & \gate{\sqrt{X}} & \ctrl{1} & \gate{\mathcal{R}_{\mathrm{z}}(\frac{\pi}{2})} & \gate{\sqrt{X}} & \ctrl{-3} & \qw & \qw & \qw & \qw & \qw & \qw & \qw & \qw & \qw & \qw & \qw & \qw & \qw & \qw & \qw & \qw & \qw & \qw & \meter \\
			\lstick{s_2 \rightarrow 7} & \ctrl{-1} & \gate{\sqrt{X}} & \gate{\mathcal{R}_{\mathrm{z}}(-\pi)} & \qw & \qw & \targ & \gate{\mathcal{R}_{\mathrm{z}}(\frac{\pi}{2})} & \qw & \qw & \targ & \gate{\mathcal{R}_{\mathrm{z}}(\frac{\pi}{2})} & \gate{\sqrt{X}} & \qw & \qw & \qw & \qw & \qw & \qw & \qw & \qw & \qw & \qw & \qw & \qw & \qw & \qw & \qw & \qw & \qw & \qw & \qw & \qw
		}
	}
	\caption{Optimum quantum circuit for \IBMQ implemeting (a) single-step gearbox including uniformly-controlled rotations \cite{mottonen_2004_generalMQGates, shende_2006_synthQLogicCircs}, (b) the simplest version of the $\Delta^\lambda$ subcircuit family, here explicitly shown for $\Delta^2$. Note that (a) corresponds to the circuit shown in Figure~6b 
	in the main text for the state register in $\ket{00}$, but also the most basic version of the $\Delta^0$ subcircuit, expected for the Hadamard gates applied on the state qubits (not shown here). The corresponding quantum circuit for (b) is depicted in Figure~9c 
	in the main text. The first columns indicate the respective virtual-to-physical qubit mapping.} 
	\label{fig::exp_circ_sgb_states}
\end{sidewaysfigure}

\FloatBarrier

\end{document}
\newpage

\documentclass[../main.tex]{subfiles}
\graphicspath{{images/}{../images/}{../images/SI images}}%
\begin{document}

\section{Quantum-Circuit Variations for the Toffoli Gate}\label{SI:toffoli}

\begin{figure}[h!]
	\begin{center}
		\includegraphics[width=0.7\textwidth]{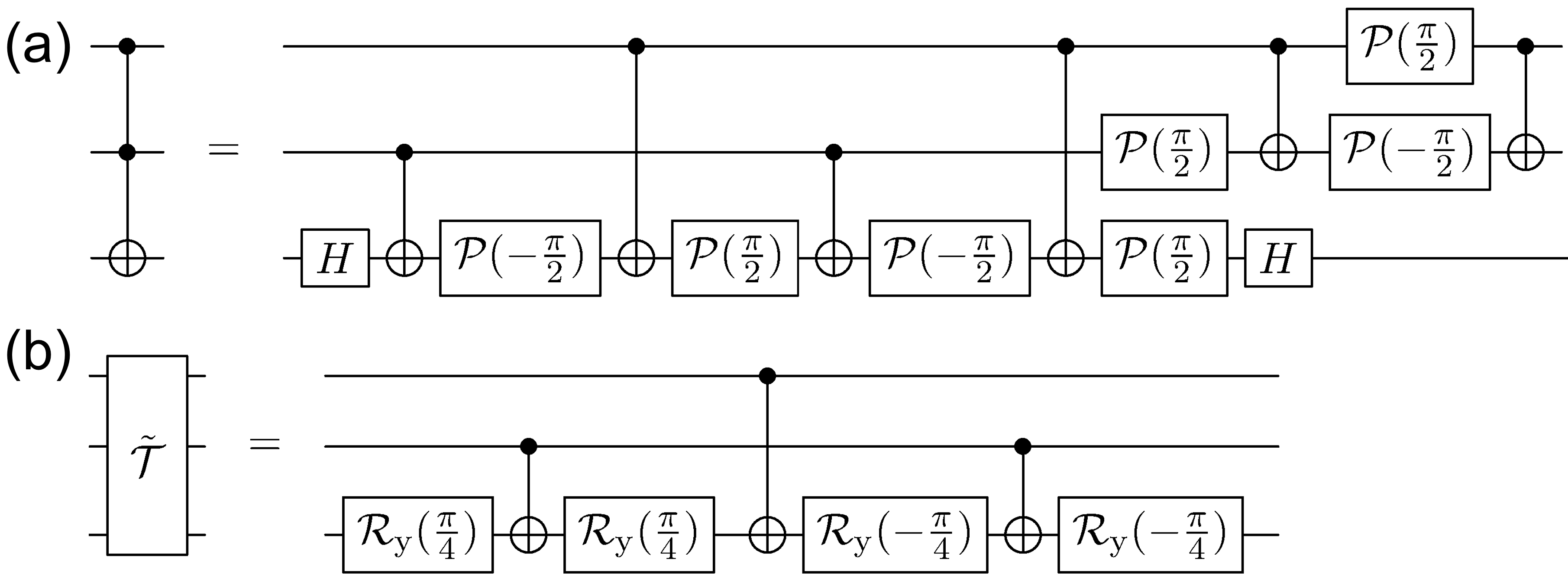}
		\caption{Quantum circuits for implementing the three-qubit Toffoli gate using single- and double-qubit gates. (a) Most efficient implementation as introduced in reference \cite{nielsen_chuang_2010, Shende_OnTC_2009}. (b) Phase-equivalent version as suggested in reference \cite{DiVincenzo_ResultsOT_1994, Barenco_elemGates_1995}.
		}
	\end{center}
	\label{fig:toffoli_def}
\end{figure}

\FloatBarrier

\end{document}
\newpage

\documentclass[../main.tex]{subfiles}
\graphicspath{{images/}{../images/}{../images/SI images}}%
\begin{document}

\section{Amplitude Subtraction of Functions}\label{SI:subtraction}

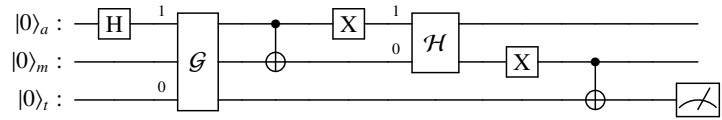
\begin{figure}[h!]
		\centering
		\scalebox{1.0}{
			\Qcircuit @C=1.0em @R=0.2em @!R {
			    \nghost{ {a}_{0} :  } & \lstick{ \ket{0}_a :  } & \gate{\mathrm{H}} & \qw & \multigate{2}{\mathrm{\mathcal{G}}}_<<<{1} & \qw & \ctrl{1} & \qw & \gate{\mathrm{X}} & \qw & \multigate{1}{\mathrm{\mathcal{H}}}_<<<{1} & \qw & \qw & \qw & \qw & \qw & \qw & \qw \\ 
				\nghost{ {m}_{0} :  } & \lstick{ \ket{0}_m :  } & \qw & \qw & \ghost{\mathrm{\mathcal{G}}} & \qw & \targ & \qw & \qw & \qw & \ghost{\mathrm{\mathcal{H}}}_<<<{0} & \qw & \gate{\mathrm{X}} & \qw & \ctrl{1} & \qw & \qw & \qw \\ 
				\nghost{ {t}_{0} :  } & \lstick{ \ket{0}_t :  } & \qw & \qw & \ghost{\mathrm{\mathcal{G}}}_<<<{0} & \qw & \qw & \qw & \qw & \qw & \qw & \qw & \qw & \qw & \targ & \qw & \qw & \meter \\ 
			}
		}
		\caption{Example of a circuit calculating the difference of two functions $g, h$ implemented by the unitary transformations $\mathcal{G}, \mathcal{H}$.}
		\label{circ:subtraction}
	\end{figure}

	\noindent \textbf{Construction.} 
	Using a modified version of the circuit that implements the addition of two functions \cite{vazquez_efficient_2020}, we can likewise calculate the difference of two functions saved as amplitudes. Even though the binomial formula enables adding two functions, we can use the counter probability to accordingly construct the subtraction. 
	%
	 Hence we may implement amplitude subtraction by the use of an additional qubit, as well as two X and two CX gates. Mathematically, the quantum circuit shown in Figure \ref{circ:subtraction} corresponds to the following calculations. Note that the indices do not indicate the length of the register, but are given to identify the corresponding register. The functions $g$ and $h$ are implemented by the unitary transformations $\mathcal{G}$ and $\mathcal{H}$, respectively. Gate applications are demonstrated above transformation arrows, where the parentheses show on which qubits the respective gates act:
	\begin{align*}
		&\ket{0}_t \ket{0}_m \ket{0}_a \\
		\overset{\mathrm{H}(a)}{\longmapsto} & \ket{0}_t \ket{0}_m \ket{+}_a \\
		\overset{\mathrm{CR}_y(a;t)}{\longmapsto} & \frac{1}{\sqrt{2}} \left[ \ket{0}_t \ket{0}_m \ket{0}_a + \left( \sqrt{1-g} \ket{0}_t + \sqrt{g} \ket{1}_t \right) \ket{0}_m \ket{1}_a \right] \\
		\overset{\mathrm{CX}(a;m)}{\longmapsto} & \frac{1}{\sqrt{2}} \left[ \ket{0}_t \ket{0}_m \ket{0}_a + \left( \sqrt{1-g} \ket{0}_t + \sqrt{g} \ket{1}_t \right) \ket{1}_m \ket{1}_a \right] \\
		\overset{\mathrm{X(a)}}{\longmapsto} & \frac{1}{\sqrt{2}} \left[ \left( \sqrt{1-g} \ket{0}_t + \sqrt{g} \ket{1}_t \right) \ket{1}_m \ket{0}_a + \ket{0}_t \ket{0}_m \ket{1}_a \right] \\
		\overset{\mathrm{CR}_y(a;m)}{\longmapsto} & \frac{1}{\sqrt{2}} \left[ \left( \sqrt{1-g} \ket{0}_t + \sqrt{g} \ket{1}_t \right) \ket{1}_m \ket{0}_a + \ket{0}_t \left( \sqrt{1-h} \ket{0}_m + \sqrt{h} \ket{1}_m \right) \ket{1}_a \right] \\
		\overset{\mathrm{X(m)}}{\longmapsto} & \frac{1}{\sqrt{2}} \left[ \left( \sqrt{1-g} \ket{0}_t + \sqrt{g} \ket{1}_t \right) \ket{0}_m \ket{0}_a + \ket{0}_t \left( \sqrt{h} \ket{0}_m + \sqrt{1-h} \ket{1}_m \right) \ket{1}_a \right] \\
		\overset{\mathrm{CX}(m;t)}{\longmapsto} & \frac{1}{\sqrt{2}} \left[ \left( \sqrt{1-g} \ket{0}_t + \sqrt{g} \ket{1}_t \right) \ket{0}_m \ket{0}_a + \left( \sqrt{h} \ket{0}_t \ket{0}_m + \sqrt{1-h} \ket{1}_t \ket{1}_m \right) \ket{1}_a \right].
	\end{align*}
	Accordingly, we get the following probability for measuring the target qubit $t$ in state $\ket{1}_t$:
	\begin{align*}
		R_1 := \frac{1}{2} \left( g + (1 - h) \right),
	\end{align*}
	and can calculate the difference by rescaling the result as
	\begin{align*}
		g - h = 2 R_1 - 1.
	\end{align*}

	\noindent \textbf{Composition.}
	Assume we want to multiply the implemented subtraction with a different unknown amplitude $z$. When combing the two qubits using a Toffoli gate with a new target qubit $q$, the resulting probability for measuring $q$ in state $\ket{1}_q$ is given by
	\begin{equation*}
	    z \cdot \frac{1}{2} ( g - h + 1 )
	    =
	    \frac{1}{2} ( z(g-h) + z ) =: b,
	\end{equation*}
	preventing to retrieve $z(g-h)$ due to the constant term $1/2$, generated by the amplitude-subtraction technique. Accordingly, an additional amplitude subtraction must be included into the circuit that implements subtracting $1/2 \cdot z$ from $b$, leading to a final probability of
	\begin{equation*}
	    \frac{1}{2} ( b - \frac{1}{2} z + 1)
	    =
	    \frac{1}{2} ( \frac{1}{2} ( z(g-h) + z ) - \frac{1}{2} z + 1)
	    =
	    \frac{1}{4} ( z(g-h) ) + \frac{1}{2} =: R_2.
	\end{equation*}
    With that we are able to obtain $z(g-h)$ from the measured probability $R_2$ by rescaling appropriately. A circuit employing this method can be seen in Figure~\ref{circ:fourier_noSplitDenom}.

\end{document}
\newpage

\documentclass[../main.tex]{subfiles}
\graphicspath{{images/}{../images/}{../images/SI images}}%
\begin{document}

\section{Implementing Arbitrary Functions as Amplitudes using Fourier Series}

	\noindent In this section, we present a method to compute any function depending on some angle $\theta$ by implementing its approximation using Fourier series expansion. Let $s(x)$ be a real-valued function with period $P$, integrable on the interval $\left[ 0, \pi/2\right]$. Note that rescaling of $s(x)$ may be required to guarantee normalization, i.e., $\tilde{s}(x)\in \left[ 0, 1\right]$. Fourier analysis states that $s(x)$ may be approximated for specific $a_i, b_i \in \mathbb{R}$ by
	\begin{equation*}
	\begin{split}
		s_N(x) = \frac{a_0}{2} + \sum_{n=1}^{N} \left( a_n \cos \left( \frac{2 \pi n}{P} x \right) + b_n \sin \left( \frac{2 \pi n}{P} x \right) \right),
	\end{split}
	\end{equation*}
	where $s_N(x) \xrightarrow{N \to \infty} s(x)$. Using $\sin(x + \pi/2) = \cos(x)$, we can rewrite $s_N$ to get
	\begin{equation*}
	\begin{split}
		s_N(x) = \frac{a_0}{2} + \sum_{n=1}^{N} \left( a_n \cos \left( \frac{2 \pi n}{P} x \right) + b_n \cos \left( \frac{2 \pi n}{P} x + \frac{\pi}{2} \right) \right).
	\end{split}
	\end{equation*}
	The previous step is not necessarily needed, but might enable future optimization techniques like possibly using the same qubit for both $\cos^2(\cdot)$ terms, where the phase shift by $\pi/4$ could be cleverly constructed using single-qubit gates.
	
	Using the double angle formula $\cos(2x) = 2\cos^2(x) - 1$, we can rewrite the above equation as
	\begin{equation*}
		\begin{split}
			s_N(x)
			&= \frac{a_0}{2} + \sum_{n=1}^{N} \left( a_n \left(2 \cos^2 \left( \frac{\pi n}{P} x \right) - 1\right) + b_n \left(2 \cos^2 \left( \frac{\pi n}{P} x + \frac{\pi}{4} \right) - 1 \right) \right) \\
			&= \underbrace{\frac{a_0}{2} - \sum_{n=1}^{N} \left( a_n + b_n \right)}_{=: a'_0} + \sum_{n=1}^{N} \left( \underbrace{2a_n}_{=: a'_n} \cos^2\left( \frac{\pi n}{P} x\right) + \underbrace{2b_n}_{=: b'_n} \cos^2\left( \frac{\pi n}{P} x + \frac{\pi}{4}\right) \right) \\
			&= a'_0 + \sum_{n=1}^{N} \left( a'_n \cos^2\left( \frac{\pi n}{P} x\right) + b'_n \cos^2\left( \frac{\pi n}{P} x + \frac{\pi}{4}\right) \right).
		\end{split}
	\end{equation*}
	This Fourier approximation can be implemented as a combination of $\mathcal{R}_\mathrm{y}$ gates using amplitude addition \cite{vazquez_efficient_2020} and subtraction, as described in Section~\ref{SI:subtraction}. The corresponding circuit is shown in Figure~\ref{circ:fourier_noSplitDenom}, the combination with a single-step gearbox circuit receiving a quantum-state input is presented in Figure~\ref{circ:complete_noSplit}. Due to its complexity, we may decrease our circuit size using the splitting method described in the main text to reliably implement it on NISQ devices \cite{preskill_2018_quantumcomputingin}. Note that in main text we have allocated each Fourier term to its own circuit, there exist alternative strategies, e.g., the separation into a positive and negative part ($\Delta(x) = \Delta_+(x) - \Delta_-(x), \Delta_+,\Delta_- \geq 0$) to avoid the amplitude subtraction procedure. The corresponding circuit can be found in Figure~\ref{circ:fourier_twoSplitDenom}.
    
\end{document}
\newpage

\documentclass[../main.tex]{subfiles}
\graphicspath{{images/}{../images/}}%
\begin{document}
\begin{sidewaysfigure}[tb]
\centering
(a)
\scalebox{.8}{
\Qcircuit @C=1.0em @R=0.2em @!R {
 	\nghost{ {s}_{0} :  } & \lstick{ {s}_{0} :  } & \qw & \multigate{2}{\sin(4\theta_x)}_<<<{2} & \qw & \multigate{3}{ \sin(2\theta_x) }_<<<{2} & \qw & \multigate{4}{ \sin(6\theta_x) }_<<<{2} & \qw & \qw & \qw & \qw & \qw & \qw & \qw & \qw & \qw & \qw & \qw & \qw & \qw\\ 
 	\nghost{ {s}_{1} :  } & \lstick{ {s}_{1} :  } & \qw & \ghost{ \sin(4\theta_x) }_<<<{1} & \qw & \ghost{ \sin(2\theta_x) }_<<<{1} & \qw & \ghost{ \sin(6\theta_x) }_<<<{1} & \qw & \qw & \qw & \qw & \qw & \qw & \qw & \qw & \qw & \qw & \qw & \qw & \qw\\ 
 	\nghost{ {u}_{0} :  } & \lstick{ {u}_{0} :  } & \qw & \ghost{ \sin(4\theta_x) }_<<<{0} & \qw & \ghost{ \sin(2\theta_x) } & \qw & \ghost{ \sin(6\theta_x) } \barrier[0em]{2} & \qw & \gate{\mathrm{X}} \barrier[0em]{2} & \qw & \ctrl{4} & \qw & \qw & \qw & \qw & \qw & \qw & \qw & \qw & \qw\\ 
 	\nghost{ {u}_{1} :  } & \lstick{ {u}_{1} :  } & \qw & \qw & \qw & \ghost{ \sin(2\theta_x) }_<<<{0} & \qw & \ghost{ \sin(6\theta_x) } & \qw & \gate{\mathrm{X}} & \qw & \qw & \qw & \qw & \ctrl{4} & \qw & \qw & \qw & \qw & \qw & \qw\\ 
 	\nghost{ {u}_{2} :  } & \lstick{ {u}_{2} :  } & \qw & \qw & \qw & \qw & \qw & \ghost{ \sin(6\theta_x) }_<<<{0} & \qw & \gate{\mathrm{X}} & \qw & \qw & \qw & \qw & \qw & \qw & \ctrl{4} & \qw & \qw & \qw & \qw\\ 
 	\nghost{ {u}_{3} :  } & \lstick{ {u}_{3} :  } & \gate{\mathrm{R_Y}\,(\mathrm{2.551})} & \ctrl{4} & \qw & \qw & \qw & \qw & \qw & \qw & \qw & \qw & \qw & \qw & \qw & \qw & \qw & \qw & \qw & \qw & \qw\\ 
 	\nghost{ {u}_{4} :  } & \lstick{ {u}_{4} :  } & \gate{\mathrm{R_Y}\,(\mathrm{0.5854})} & \qw & \qw & \qw & \qw & \qw & \qw & \qw & \qw & \ctrl{3} & \qw & \qw & \qw & \qw & \qw & \qw & \qw & \qw & \qw\\ 
 	\nghost{ {u}_{5} :  } & \lstick{ {u}_{5} :  } & \gate{\mathrm{R_Y}\,(\mathrm{1.541})} & \qw & \qw & \qw & \qw & \qw & \qw & \qw & \qw & \qw & \qw & \qw & \ctrl{3} & \qw & \qw & \qw & \qw & \qw & \qw\\ 
 	\nghost{ {u}_{6} :  } & \lstick{ {u}_{6} :  } & \gate{\mathrm{R_Y}\,(\mathrm{0.2396})} & \qw & \qw & \qw & \qw & \qw & \qw & \qw & \qw & \qw & \qw & \qw & \qw & \qw & \ctrl{2} & \qw & \qw & \qw & \qw\\ 
 	\nghost{ {a}_{0} :  } & \lstick{ {a}_{0} :  } & \gate{\mathrm{H}} & \ctrl{2} & \qw & \gate{\mathrm{X}} & \qw & \qw & \qw & \qw & \qw & \ctrl{2} & \qw & \qw & \qw & \qw & \qw & \qw & \qw & \qw & \qw\\ 
 	\nghost{ {a}_{1} :  } & \lstick{ {a}_{1} :  } & \gate{\mathrm{H}} & \qw & \qw & \qw & \qw & \qw & \qw & \qw & \qw & \qw & \qw & \qw & \ctrl{1} & \gate{\mathrm{X}} & \ctrl{1} & \qw & \qw & \qw & \qw\\ 
 	\nghost{ {a}_{2} :  } & \lstick{ {a}_{2} :  } & \gate{\mathrm{H}} & \ctrl{2} & \qw & \qw & \qw & \qw & \qw & \qw & \qw & \ctrl{2} & \ctrl{1} & \gate{\mathrm{X}} & \ctrl{1} & \qw & \ctrl{1} & \qw & \qw & \qw & \qw\\ 
 	\nghost{ {a}_{3} :  } & \lstick{ {a}_{3} :  } & \qw & \qw & \qw & \qw & \qw & \qw & \qw & \qw & \qw & \qw & \targ & \qw & \targ & \qw & \targ & \gate{\mathrm{X}} & \ctrl{1} & \qw & \qw\\ 
 	\nghost{ {d}_{0} :  } & \lstick{ {d}_{0} :  } & \qw & \targ & \qw & \qw & \qw & \qw & \qw & \qw & \qw & \targ & \qw & \qw & \qw & \qw & \qw & \qw & \targ & \qw & \qw\\ 
    }
}\newline
\newline%
(b)
\scalebox{.8}{ 
\Qcircuit @C=1.0em @R=0.2em @!R {
    \nghost{ {z}_{0} :  } & \lstick{ {z}_{0} :  } & \multigate{2}{z(\theta_x)}_<<<{0} & \qw & \ctrl{12} & \qw & \qw & \ctrl{11} & \qw & \qw & \qw & \qw \\
 	\nghost{ {s}_{0} :  } & \lstick{ {s}_{0} :  } & \ghost{z(\theta_x)}_<<<{1} & \multigate{7}{d(\theta_x)}_<<<{3} & \qw & \qw & \qw & \qw & \qw & \qw & \qw & \qw \\
 	\nghost{ {s}_{1} :  } & \lstick{ {s}_{1} :  } & \ghost{z(\theta_x)}_<<<{2} & \ghost{d(\theta_x)}_<<<{2} & \qw & \qw & \qw & \qw & \qw & \qw & \qw & \qw \\
 	\nghost{ {u}_{0} :  } & \lstick{ {u}_{0} :  } & \qw & \ghost{d(\theta_x)}_<<<{1} & \qw & \qw & \qw & \qw & \qw & \qw & \qw & \qw \\ 
 	& & & & & & & & & &\\
	\hspace{1.3cm} \vdots & & & & & & & & & &\\
	& & & & & & & & & &\\
 	\nghost{ {a}_{3} :  } & \lstick{ {a}_{3} :  } & \qw & \ghost{d(\theta_x)}_<<<{1} & \qw & \qw & \qw & \qw & \qw & \qw & \qw & \qw \\ 
 	\nghost{ {d}_{0} :  } & \lstick{ {d}_{0} :  } & \qw & \ghost{d(\theta_x)}_<<<{0} & \ctrl{1} & \qw & \qw & \qw & \qw & \qw & \qw & \qw \\ 
    \nghost{ {a}_{4} :  } & \lstick{ {a}_{0} :  } & \gate{\mathrm{H}} & \qw & \ctrl{1} & \ctrl{2} & \gate{\mathrm{X}} & \ctrl{1} & \qw & \qw & \qw & \qw \\ 
    \nghost{ {a}_{5} :  } & \lstick{ {a}_{1} :  } & \gate{\mathrm{H}} & \qw & \qw & \qw & \qw & \ctrl{1} & \qw & \qw & \qw & \qw \\ 
    \nghost{ {m}_{0} :  } & \lstick{ {m}_{0} :  } & \qw & \qw & \qw & \targ & \qw & \targ & \gate{\mathrm{X}} & \ctrl{1} & \qw & \qw \\
    \nghost{ {f}_{0} :  } & \lstick{ {f}_{0} :  } & \qw & \qw & \targ & \qw & \qw & \qw & \qw & \targ & \qw & \meter\\
    }
}%
\caption{Direct implementation of the Fourier approximation $D(x)/2 = 0.91526 + 0.0832611 \cos(4x)^2 - 0.485281 \cos(2x)^2 - 0.0142853 \cos(6x)^2$ of $\rho^{-2}(\theta)$ using Amplitude Subtraction. (a) Implementing the Fourier approximation $d(\theta_x)$, which needs 142 CX-gate applications on the quantum simulator from Qiskit. (b) Combining $D(\theta_x)$ with a different function $z(\theta_x)$, which needs an additional 74 CX-gate applications.}  
\label{circ:fourier_noSplitDenom}
\end{sidewaysfigure}
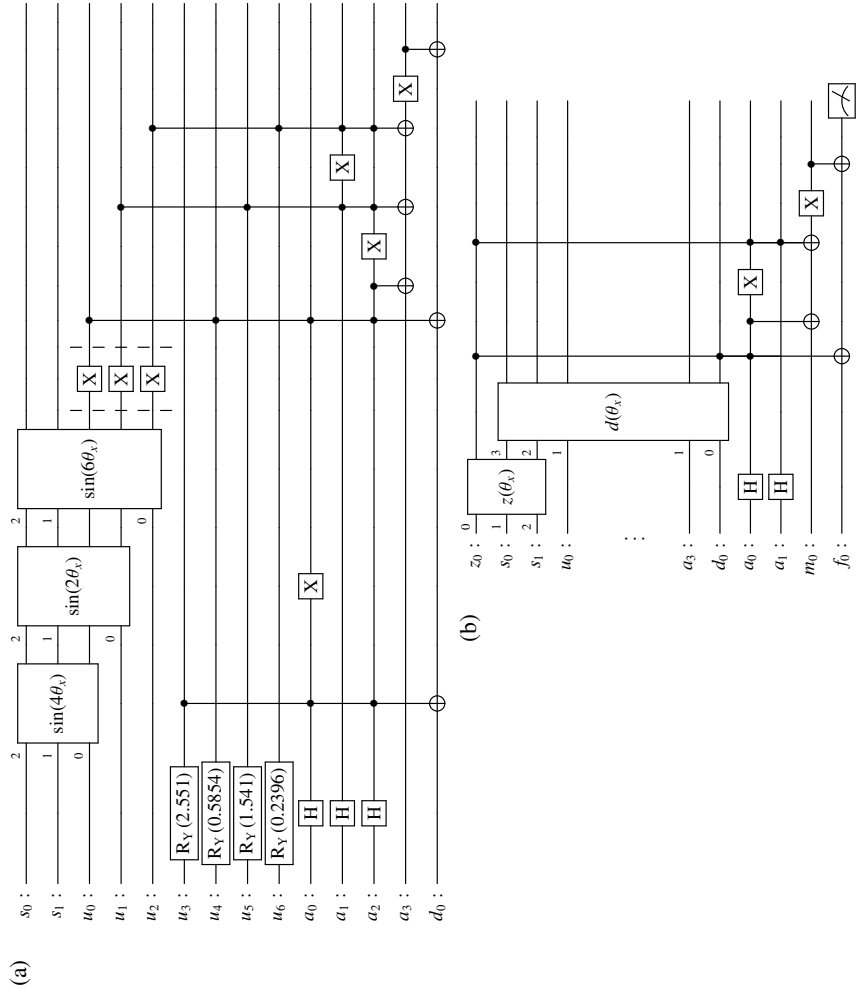
\end{document}

\newpage


\documentclass[../main.tex]{subfiles}
\graphicspath{{images/}{../images/}}%
\begin{document}
\begin{figure}[tb]
\centering
\scalebox{.8}{
    \Qcircuit @C=1.0em @R=0.2em @!R {
	 	\nghost{{s}_{0} :  } & \lstick{{s}_{0} :  } & \gate{\mathrm{H}} & \multigate{4}{\Gb(\pmb{\theta})}_<<<{4} & \qw & \qw & \qw & \qw & \multigate{11}{D(\pmb{\theta})}_<<<{13} & \qw \\
	 	\nghost{{s}_{1} :  } & \lstick{{s}_{1} :  } & \gate{\mathrm{H}} & \ghost{\Gb(\pmb{\theta})}_<<<{3} & \qw & \qw & \qw & \qw & \ghost{D(\pmb{\theta})}_<<<{12} & \qw \\
	 	\nghost{{c}_{0} :  } & \lstick{{c}_{0} :  } & \qw & \ghost{\Gb(\pmb{\theta})}_<<<{2} & \qw & \qw & \qw & \qw & \ghost{D(\pmb{\theta})} & \qw \\
	 	\nghost{{c}_{1} :  } & \lstick{{c}_{1} :  } & \qw & \ghost{\Gb(\pmb{\theta})}_<<<{1} & \qw & \qw & \qw & \qw & \ghost{D(\pmb{\theta})} & \qw \\
	 	\nghost{{t}_{0} :  } & \lstick{{t}_{0} :  } & \qw & \ghost{\Gb(\pmb{\theta})}_<<<{0} & \ctrl{1} & \gate{\mathrm{X}} & \ctrl{1} & \gate{\mathrm{X}} & \ghost{D(\pmb{\theta})} & \qw \\
	 	\nghost{{o}_{0} :  } & \lstick{{o}_{0} :  } & \qw & \qw & \targ & \qw & \gate{\mathcal{R}_\mathrm{Y}(x)} & \qw & \ghost{D(\pmb{\theta})} & \qw \\
	 	\nghost{{a_{0}} :  } & \lstick{{a}_{0} :  } & \qw & \qw & \qw & \qw & \qw & \qw & \ghost{D(\pmb{\theta})}_<<<{11} & \qw \\
    	& & & & & & & & & &\\
    	\hspace{1.3cm} \vdots & & & & & \hspace{1.3cm} \vdots & & & & &\\
    	& & & & & & & & & &\\
    	\nghost{{a_{10}} :  } & \lstick{{a_{10}} :  } & \qw & \qw & \qw & \qw & \qw & \qw & \ghost{D(\pmb{\theta})}_<<<<{1} & \qw \\
	    \nghost{{f}_{0} :  } & \lstick{{f}_{0} :  } & \qw & \qw & \qw & \qw & \qw & \qw & \ghost{D(\pmb{\theta})}_<<<<{0} & \meter \\
    }
}
\caption{Our implementation of $\Omega(10)$ using the non-splitted $\Delta$ is presented. The measured qubit is found in state $\ket{1}_f$ with a probability of approximately $(\overline{\mathcal{S}^{\circ 1}} / 16 ) + 1/2 $. The quantum circuit uses 
$238$ CX gates on the quantum simulator from Qiskit, because of which there is no possibility of currently implementing it on quantum hardware.}
\label{circ:complete_noSplit}
\end{figure}
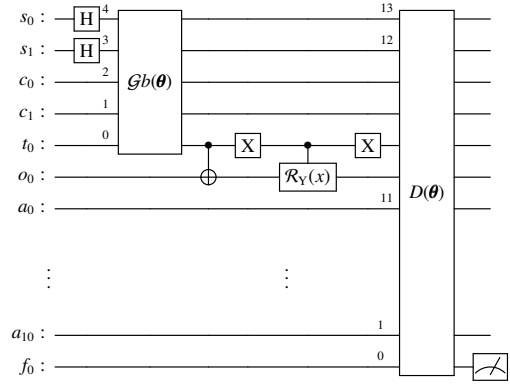
\end{document}

\newpage

\documentclass[../main.tex]{subfiles}
\graphicspath{{images/}{../images/}}%
\begin{document}
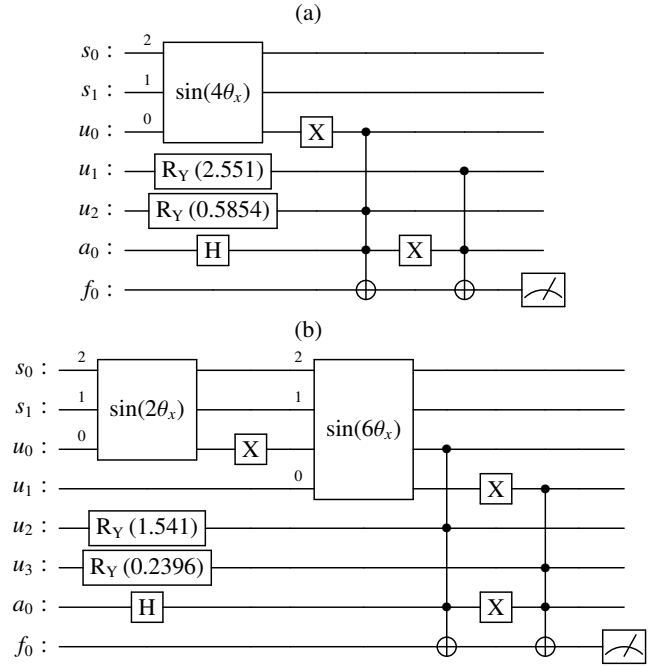
\begin{figure}[tb]
\centering
(a)

\hspace{2.5cm}
\scalebox{1.0}{
    \Qcircuit @C=1.0em @R=0.2em @!R {
	 	\nghost{ {s}_{0} :  } & \lstick{ {s}_{0} :  } & \multigate{2}{\sin(4\theta_x)}_<<<{2} & \qw & \qw & \qw & \qw & \qw & \qw\\ 
	 	\nghost{ {s}_{1} :  } & \lstick{ {s}_{1} :  } & \ghost{\sin(4\theta_x)}_<<<{1} & \qw & \qw & \qw & \qw & \qw & \qw\\ 
	 	\nghost{ {u}_{0} :  } & \lstick{ {u}_{0} :  } & \ghost{\sin(4\theta_x)}_<<<{0} & \gate{\mathrm{X}} & \ctrl{2} & \qw & \qw & \qw & \qw\\ 
	 	\nghost{ {u}_{1} :  } & \lstick{ {u}_{1} :  } & \gate{\mathrm{R_Y}\,(\mathrm{2.551})} & \qw & \qw & \qw & \ctrl{2} & \qw & \qw\\ 
	 	\nghost{ {u}_{2} :  } & \lstick{ {u}_{2} :  } & \gate{\mathrm{R_Y}\,(\mathrm{0.5854})} & \qw & \ctrl{1} & \qw & \qw & \qw & \qw\\ 
	 	\nghost{ {a}_{0} :  } & \lstick{ {a}_{0} :  } & \gate{\mathrm{H}} & \qw & \ctrl{1} & \gate{\mathrm{X}} & \ctrl{1} & \qw & \qw\\ 
	 	\nghost{ {f}_{0} :  } & \lstick{ {f}_{0} :  } & \qw & \qw & \targ & \qw & \targ & \qw & \meter\\ 
    }
}\newline
\newline%
(b)

\scalebox{1.0}{
    \Qcircuit @C=1.0em @R=0.2em @!R {
	 	\nghost{ {s}_{0} :  } & \lstick{ {s}_{0} :  } & \multigate{2}{\sin(2\theta_x)}_<<<{2} & \qw & \qw & \multigate{3}{\sin(6\theta_x)}_<<<{2} & \qw & \qw & \qw & \qw & \qw\\ 
	 	\nghost{ {s}_{1} :  } & \lstick{ {s}_{1} :  } & \ghost{\sin(2\theta_x)}_<<<{1} & \qw & \qw & \ghost{\sin(6\theta_x)}_<<<{1} & \qw & \qw & \qw & \qw & \qw\\ 
	 	\nghost{ {u}_{0} :  } & \lstick{ {u}_{0} :  } & \ghost{\sin(2\theta_x)}_<<<{0} & \gate{\mathrm{X}} & \qw & \ghost{\sin(6\theta_x)} & \ctrl{2} & \qw & \qw & \qw & \qw\\ 
	 	\nghost{ {u}_{1} :  } & \lstick{ {u}_{1} :  } & \qw & \qw & \qw & \ghost{\sin(6\theta_x)}_<<<{0} & \qw & \gate{\mathrm{X}} & \ctrl{2} & \qw & \qw\\ 
	 	\nghost{ {u}_{2} :  } & \lstick{ {u}_{2} :  } & \gate{\mathrm{R_Y}\,(\mathrm{1.541})} & \qw & \qw & \qw & \ctrl{2} & \qw & \qw & \qw & \qw\\ 
	 	\nghost{ {u}_{3} :  } & \lstick{ {u}_{3} :  } & \gate{\mathrm{R_Y}\,(\mathrm{0.2396})} & \qw & \qw & \qw & \qw & \qw & \ctrl{1} & \qw & \qw\\ 
	 	\nghost{ {a}_{0} :  } & \lstick{ {a}_{0} :  } & \gate{\mathrm{H}} & \qw & \qw & \qw & \ctrl{1} & \gate{\mathrm{X}} & \ctrl{1} & \qw & \qw\\ 
	 	\nghost{ {f}_{0} :  } & \lstick{ {f}_{0} :  } & \qw & \qw & \qw & \qw & \targ & \qw & \targ & \qw & \meter\\ 
    }
}%
\caption{Direct implementation of the Fourier approximation $D(x)=D_+(x)+D_-(x)$ of $\rho^{-2}(\theta)$ for two state qubits using amplitude addition. (a) Implementing the positive parts of the Fourier approximation $ D_+(x)/2 = 0.91526 + 0.0832611 \cos(4x)^2 $ on a quantum circuit. (b) Implementing the negative parts of the Fourier approximation $ D_-(x)/2 = 0.485281 \cos(2x)^2 + 0.0142853 \cos(6x)^2$ on a quantum circuit.}
\label{circ:fourier_twoSplitDenom}
\end{figure}
\end{document}

\newpage

\end{document}

\FloatBarrier

\bibliographystyle{naturemag}
\bibliography{lit}


\section{Splitting the Fourier approximation}

	As a starting point, we need one circuit for the positive, and one for the negative part of the Fourier approximation $D(x)$. We can do this because for some given circuit with probability of the result qubit to be in state $\ket{1}$ of $z(x)$ coupled with $D(x)$ using a Toffoli gate, we get a resulting probability of our new target qubit being $\ket{1}$ of
	\begin{equation*}
	\begin{split}
		\sum_{k=0}^{2^N-1} z(x_k) D(x_k)
		&= \sum_{k=0}^{2^N-1} z(x_k) \left(D^+(x_k) - D^-(x_k)\right) \\
		&= \underbrace{\sum_{k=0}^{2^N-1} z(x_k) D^+(x_k) }_{=\text{circuit 1}} - \underbrace{\sum_{k=0}^{2^N-1} z(x_k) D^-(x_k)}_{=\text{circuit 2}}.
	\end{split}
	\end{equation*}
	You can see that we may split the Fourier approximation into positive and negative parts $D(x) = D^+(x) - D^-(x), D^+(x), D^-(x) \geq 0$. Keep in mind that this split of the approximation into two parts is not unique. If we now implement circuit~1 and circuit~2 and measure their results independently, subtracting the second measurement from the first, we get the wanted result. The corresponding quantum circuit is shown in Figure~\ref{circ:fourier_twoSplitDenom}. The `optimal' splitting may be object to future research, as it strongly depends on the defined objective and problem size at hand.